\documentclass[conference]{IEEEtran}

\ifCLASSINFOpdf
\else
\fi

\PassOptionsToPackage{hyphens}{url}
\usepackage[breaklinks]{hyperref}
\usepackage{amssymb}
\usepackage{longtable}
\usepackage{booktabs}
\usepackage{graphicx}
\usepackage{tikz} 
\usepackage{tabularx}
\usepackage{fp}
\usepackage{array}
\usepackage{svg}
\usepackage{amsmath}
\usepackage{adjustbox}
\usepackage{colortbl}
\usepackage{tcolorbox}
\usepackage{arydshln}
\usepackage{pdflscape}
\usepackage{enumitem}
\usepackage{siunitx}
\usepackage{cleveref}
\usepackage{fontawesome}
\usepackage{xcolor}
\definecolor{munsell}{rgb}{0.0, 0.5, 0.69}
\usepackage[edges]{forest}
\usepackage[framemethod=TikZ]{mdframed}
\usepackage{multirow}
\usepackage{threeparttable}
\usepackage{multicol}
\usepackage{wasysym}
\usepackage{graphicx}
\usepackage{makecell}
\usepackage{pifont}
\usepackage{stix}
\usepackage{listings}

\newcommand{\empirical}[1]{#1}

\definecolor{Colorset1}{HTML}{F05039}
\definecolor{Colorset2}{HTML}{E57A77}
\definecolor{Colorset3}{HTML}{EEBAB4}
\definecolor{Colorset4}{HTML}{1F449C}
\definecolor{Colorset5}{HTML}{3D65A5}
\definecolor{Colorset6}{HTML}{7CA1CC}
\definecolor{Colorset7}{HTML}{A8B6CC}

\newlength{\mdinnermarginlen}
\newmdenv [%
    roundcorner=2.5pt
    outerlinewidth=0pt,
    innerlinewidth=0pt,
    backgroundcolor=black!10,
    outerlinecolor=black,
    linecolor=black,
    nobreak=false,
    innerleftmargin=\mdinnermarginlen,
    innerrightmargin=\mdinnermarginlen,
    innertopmargin=\mdinnermarginlen,
    innerbottommargin=\mdinnermarginlen,
]{defenv}

\newcounter{defcntr}

\newmdenv [%
    roundcorner=2.5pt
    outerlinewidth=0pt,
    innerlinewidth=0pt,
    backgroundcolor=black!10,
    outerlinecolor=black,
    linecolor=black,
    nobreak=false,
    innerleftmargin=\mdinnermarginlen,
    innerrightmargin=\mdinnermarginlen,
    innertopmargin=\mdinnermarginlen,
    innerbottommargin=\mdinnermarginlen,
]{taxenv}
\newcommand{\taxonomy}[1]{%
\vspace{1ex}
\begin{taxenv}%
\small\faEye\,\textbf{Summary:} #1
\end{taxenv}%
}

\newcommand{\keyinsight}[1]{%
\vspace{1ex}
\begin{taxenv}%
\small\faEye\,\textbf{Key Insight:} #1
\end{taxenv}%
}

\newcolumntype{Y}{>{\centering\arraybackslash}X}
\newcommand{\faDoubleCheck}{\faCheck\kern-0.5em\faCheck}

\newcommand{\sdot}[1]{%
  \textsuperscript{\tikz \draw[#1, fill=#1, radius=2pt] (0,0) circle ;}%
}

\definecolor{cbcdxgen}{HTML}{004488}
\definecolor{cbsyft}{HTML}{DDAA33}
\definecolor{cbtrivy}{HTML}{117733}
\definecolor{cbort}{HTML}{882255}
\definecolor{cbsbomtool}{HTML}{44AA99}

\newcommand{\filescdxgen}[0]{\sdot{cbcdxgen}}
\newcommand{\filessyft}[0]{\sdot{cbsyft}}
\newcommand{\filestrivy}[0]{\sdot{cbtrivy}}
\newcommand{\filesort}[0]{\sdot{cbort}}
\newcommand{\filessbomtool}[0]{\sdot{cbsbomtool}}

\newcommand{\cdxgen}[0]{cdxgen}
\newcommand{\syft}[0]{syft}
\newcommand{\trivy}[0]{trivy}
\newcommand{\ort}[0]{ORT}
\newcommand{\sbomtool}[0]{sbom-tool}

\newcommand{\Cdxgen}[0]{Cdxgen}
\newcommand{\Syft}[0]{Syft}
\newcommand{\Trivy}[0]{Trivy}
\newcommand{\Ort}[0]{ORT}
\newcommand{\SBOMTOOL}[0]{SBOM Tool}

\let\textquotedbl="

\makeatletter
\newcommand{\linebreakand}{%
  \end{@IEEEauthorhalign}
  \hfill\mbox{}\par
  \mbox{}\hfill\begin{@IEEEauthorhalign}
}
\makeatother

\begin{document}

\title{Poking Around in the Dark: Why a Shared Understanding of Components Matters}

\author{
    \IEEEauthorblockN{Felix Reichmann}
	\IEEEauthorblockA{Ruhr University Bochum\\
		\href{mailto:felix.reichmann@ruhr-uni-bochum.de}{felix.reichmann@ruhr-uni-bochum.de}}
	\and
	\IEEEauthorblockN{Wolfgang Krane}
	\IEEEauthorblockA{\href{mailto:wolfgang.krane@mausketier.de}{wolfgang.krane@mausketier.de}}
	\and
	\IEEEauthorblockN{Alena Naiakshina}
	\IEEEauthorblockA{University of Cologne\\
        \href{mailto:alena.naiakshina@uni-koeln.de}{alena.naiakshina@uni-koeln.de}}
    \linebreakand
    \IEEEauthorblockN{Martin Johns}
	\IEEEauthorblockA{TU Braunschweig\\
        \href{mailto:m.johns@tu-braunschweig.de}{m.johns@tu-braunschweig.de}}
    \and
	\IEEEauthorblockN{Simon Koch}
	\IEEEauthorblockA{University of Innsbruck\\
        \href{simon.koch@uibk.ac.at}{simon.koch@uibk.ac.at}}
}
	
\maketitle

\begin{abstract}
By listing the components included in an application, Software Bills of Materials (SBOMs) are intended to support the timely identification of vulnerable components and ensure the security of the software supply chain.
However, we question the underlying assumption that there is agreement on the components to be listed in an SBOM and that current technology is sufficient to secure the software supply chain.

First, we propose a ground-up analysis of Component Inclusion Mechanisms (CIM) in the software's development lifecycle.
Then we systematically analyze the four popular SBOM generation tools, \cdxgen{}, \syft{}, \trivy{}, \ort{}, and the Microsoft \sbomtool{}, to understand how they define and identify relevant components.
Finally, we assess these using a ground truth across the programming languages Python, Java, Go, PHP, Rust, and C.

While today's tools are a step toward identifying components, our results show that no tool covers all identified CIMs and that common gaps exist across tools.
We demonstrate that, under the current vague definitions and tooling, SBOMs exhibit ambiguity and blind spots in component inclusion.
Thus, a security-grade SBOM is not achievable with the evaluated tools, necessitating further progress to ensure software supply chain security.
We need to go back to the drawing board to clarify which components should be included in an SBOM and revise SBOM generators accordingly.
Without a shared understanding of what a component is, any effort to secure software supply chains with SBOMs will fail.
\end{abstract}

\IEEEpeerreviewmaketitle

\section{Introduction}\label{Introduction}
The Log4Shell vulnerability demonstrated how challenging it is to respond to supply chain vulnerabilities: First, a patch had to be available, and second, organizations needed to identify and apply the update to the affected products and systems~\cite{Vaas.2022, Umbelino.2022, Springett.2025, Ahmed.2024}.
While a patch was quickly available after 16 days~\cite{CISA.2022}, the remediation posed a greater challenge.
The Cyber Safety Review Board of the Cybersecurity and the Infrastructure Security Agency (CISA) suggests that the Log4Shell vulnerability could persist for ``a decade or longer''~\cite{CISA.2022}.
This is due to the complexity and nesting of software supply chains~\cite{Mead.2024}.
Other past high-impact vulnerabilities in software and their components, such as Heartbleed~\cite{Heartbleed.NVD.2014} or the recent backdoor in XZ Utils~\cite{Merigala.2024} further underscored the fragility of the software ecosystem and the need to reliably identify vulnerable components.

Regulatory bodies have recognized this need and its resulting challenges and are taking action accordingly.
For example, in the United States, Executive Order 14028 was issued to improve the software supply chain security~\cite{EO14028}, and the European Union proposed the Cyber Resilience Act (CRA)~\cite{CRA.2024}.
The Open Worldwide Application Security Project (OWASP) recognized ``Software Supply Chain Failures'' as number 3 of the 10 most critical security risks to web applications~\cite{OWASP.Top10.2025}.
However, as of today, SBOMs face two challenges:

\emph{The second challenge} is the reliability and quality of today's tools that are deployed to generate SBOMs.
High-quality SBOMs are crucial for drawing reliable conclusions, but practitioners suggest that generating accurate SBOMs is a challenge~\cite{Stalnaker.2024, Bi.2024, Zahan.2023} and existing research has shown issues with
(a) single programming languages~\cite{Xiao.2025, Balliu.2023, Rabbi.2024, Cofano.2024, Wang.2026b}, 
(b) the presence of required standardized fields~\cite{Halbritter.2024, ManziMuneza.2025, TorresArias.2023}, 
(c) diverging output between different tools~\cite{Yu.2024},
(d) SBOMs derived from container images and their impact on vulnerability detection~\cite{ManziMuneza.2025}, or
(e) the completeness of components on GitHub's dependency graph using bidirectional checks~\cite{Bifolco.2024}.

However, the \emph{first and foremost question} remains unaddressed: \emph{What should actually be counted as a software component?}
Without a clear definition, tool builders apply their own subjective interpretations of what constitutes a component.
As a result, SBOMs may differ between tools and become ambiguous rather than a clear, reliable listing.
This also poses challenges for researchers: How can we assess the completeness of an SBOM if there is no common understanding of what ``complete'' means?
While it seems trivial at first glance to define the term ``component,'' it proves challenging on second thought.
Copied code adds functionality to an application, but is not added in a structured way.
Is it a component?
Is dynamically loaded code, such as plugins, a component?
These questions are challenging since international cybersecurity agencies and SBOM standard providers have differing definitions of a component~\cite{CRA.2024, TR-03183-2.2024, SPDX.package.2024, NTIA.2021e, EO14028}.
Similarly, relevant research also seems to be inconsistent, as some work considers only managed dependencies to be components~\cite{Cofano.2024, Balliu.2023, Rabbi.2024}, while other provide a broader interpretation~\cite{Xiao.2025, Wang.2026b, Kong.2025}.

We address this gap from a security researcher's perspective and follow a three-step approach:
(1) We derive a definition for a component grounded on the overarching security aim of SBOMs and structurally identify the different mechanisms of introducing those components into a software (Section~\ref{taxonomy}).
(2) We construct an automated assessment for those inclusion mechanisms across Python, Java, Go, PHP, Rust, and C and evaluate 5 leading SBOM generation tools (\cdxgen{}~\cite{cdxgen}, \syft{}~\cite{Syft}, \trivy{}~\cite{Trivy}, \ort{}~\cite{LinuxFoundation.ORT}, \sbomtool{}~\cite{MS.SBOMTOOL}), highlighting the current blind spots in SBOMs (Section~\ref{Benchmark}).
(3) To determine if any detected blind spot is due to a lack of technical capability, we finally perform a manual code analysis (Section~\ref{sec:generatorBisection}).

In summary, our work addresses the following three research questions:
\begin{enumerate}[label=\textbf{RQ\arabic*}, leftmargin=*]
    \setlength\itemsep{0em}
    \setlength\parskip{0em}
    \setlength\parsep{0em}

    \item \textbf{What should be considered a software component?}\\ \textit{(Section~\ref{taxonomy})}

    \item \textbf{How well do different tools across different programming languages detect software components?}\\ \textit{(Section~\ref{Benchmark})}

    \item \textbf{Why do different tools only detect a subset of expected components?}\\ \textit{(Section~\ref{sec:generatorBisection})}
\end{enumerate}

We found that current SBOM generators effectively detect managed components but almost entirely miss build- and runtime-inclusion mechanisms, leaving large blind spots.
We adhere to open-source research, and our developed tooling is publicly available\footnote{pending publication}.
This encompasses the assessment suite, including code and documentation.
\section{Background \& Related Work}\label{Background}
To understand the challenge of Component Inclusion and how SBOM generators struggle with it, we first need to introduce SBOMs and the various standards (Section~\ref{rel:RoleOfSBOMs}).
Second, we examine existing research on SBOM quality (Section~\ref{rel:QualityEvaluation}).

\subsection{What is an SBOMs?}\label{rel:RoleOfSBOMs}
An SBOM is a standardized list of components and is defined as ``a formal record of the details and supply chain relationships of various components used''~\cite{EO14028, CISA.2025, NIST.Def.SBOM.2026}.
Three machine-readable formats have become widespread~\cite{Stalnaker.2024, NTIA.2021c}: Software Identification (SWID) Tagging, defined in ISO/IEC 19770-2:2015~\cite{ISO.19770.2015}, CycloneDX~\cite{CycloneDX} from OWASP, and SPDX~\cite{SPDX} from the Linux Foundation.
SPDX and CycloneDX are the two most widely used standards~\cite{Cofano.2024, Sonatype.2023}.
Common output formats are JSON or XML.

While an SBOM provides an overview of the components contained in a software, the actual added security value comes from the fact that the components can be automatically compared against a vulnerability database such as the NIST NVD~\cite{NVD, Benedetti.2025b}.
This allows vulnerable components to be efficiently identified, thereby significantly accelerating the mitigation process~\cite{NTIA.2019b, CISA.2025}.

The listing of a software product's contents can be performed at different stages of the Software Development Life Cycle (SDLC).
CISA and BSI therefore define six types of SBOMs based on the stage used~\cite{CISA.2023b, TR-03183-2.2024}: Design, Source, Build, Analyzed, Deployed, and Runtime.
\Cref{tbl:sbomtypes} in the Appendix provides a visual overview.

\subsection{Quality Evaluations}\label{rel:QualityEvaluation}
Accurate SBOMs are critical for automated vulnerability detection, as missing or misidentified components lead to false negatives or positives.
To obtain reliable SBOMs, prior work has begun evaluating SBOM quality in various ways.

Torres-Arias et al.~\cite{TorresArias.2023} used the NTIA conformity checker and the SBOM scorecard to evaluate which fields were present across generators.
Additionally, they point out that there is no baseline for which components should be documented; therefore, they perform only a differential analysis, comparing how many components Trivy and Syft found for the same project.
Citing the absence of a ground truth as well, Yu et al.~\cite{Yu.2024} performed a more extensive differential analysis.
Bifolco et al.~\cite{Bifolco.2024} analyzed the quality of the GitHub Dependency Graph~\cite{GitHubDependencyGraph} by bidirectional checking if backward and forward projects are referencing each other.

Halbritter et al.~\cite{Halbritter.2024} primarily assessed whether SBOM tools populate the structural metadata fields mandated by the NTIA Minimum Elements~\cite{NTIA.2021b}, using test applications across four languages that relied exclusively on package manager components.
As a secondary metric, they measured ``completeness'' for these managed components, providing initial results for component detection by different SBOM generators.
Similarly, Muneza et al.~\cite{ManziMuneza.2025} evaluated the influence of SBOM generation tools and formats on compliance with the NTIA standard.
The approach of evaluating SBOMs based on syntactic compliance is also supported by various industry projects.
These include industry driven projects, like the \textit{SBOM Scorecard} by eBay~\cite{Ebay.2022}, \textit{sbomqs} by Interlynk~\cite{Interlynk.2023}, \textit{\syft} by Anchore~\cite{Syft}, \textit{cyclonedx-editor-validator} by Festo~\cite{Festo.2023} or the community driven projects \textit{Lib4SBOM} and \textit{SBOMAUDIT} by Anthony Harrison~\cite{Harrison.2023, Harrison.2023b}.
Additionally, the standardization bodies provide their own tools for SPDX~\cite{SPDX.Toolings.2016, SPDX.NTIA.ComplianceChecker.2022} and CycloneDX~\cite{CycloneDX.CLI.2020}.
However, this has the clear drawback that they cannot assess whether all relevant components are listed in the SBOM.
To achieve this, two different methods have emerged for generating a ground truth and using it to evaluate the detected components:

\emph{Generating a ground truth for real-world projects.}
Balliu et al.~\cite{Balliu.2023} and Rabbi et al.~\cite{Rabbi.2024} used native package manager outputs as a ground truth.
While this method allows for a systematic evaluation of the identified components, the approach has a significant limitation: only elements listed by the package managers are considered components.
However, other studies have defined this component more comprehensively.
Xiao et al.~\cite{Xiao.2025}, who evaluated SBOMs in the Maven Central ecosystem, developed a tool to derive ground truth from the corresponding JAR files.
In doing so, they took into account not only managed components but also dynamic features, whereby these posed a greater challenge and they found that ``inefficiencies in dynamic feature resolution also led to missed detections''.
Similarly, Wang et al.~\cite{Wang.2026b} recognized the challenge of concentrating on managed components. While the main focus of their study is on the compliance and consistency of real-world SBOMs in different languages, they developed a tool that retrieves examples for Python and a ground truth from the source code, i.e., by analyzing import statements.
Kong et al.~\cite{Kong.2025} have developed an approach to identify components in source code, with particular emphasis on component accuracy.
They test their tool by identifying components in 30 open-source projects and comparing their results with the SBOMs generated by \cdxgen{}, openrewrite, build-info-go, and the Microsoft SBOM Tool.
These studies evaluate quality based on real-world projects and thus provide a good picture of the current state of affairs, but they are highly ecosystem-specific.

\emph{Synthetic test programs.}
Another approach is not to generate a ground truth for real-world projects, but rather to create synthetic test projects. Even though the number of projects evaluated is smaller in this case, it allows for very specific testing of certain capabilities of the SBOM generators.
Cofano et al.~\cite{Cofano.2024} developed a synthetic dataset for Python in which they incorporated managed components in various ways.
Similarly, Garcia et al.~\cite{Garcia.2025}, identified 84 tools in the SBOM ecosystem and evaluated 5 of them against 3 self-developed Java applications, focusing on managed components.

\textbf{Research Gap.}
No prior work has addressed what a component actually is, which is essential before assessing existing tools.
We first answer this question and then combine the two approaches presented for SBOM evaluation by providing real-world test cases with a known, well-defined ground truth.
With this work, we provide a test suite that allows manufacturers to identify and address gaps in their tools.
\section{Component Inclusion Mechanisms}\label{sec:dependencyInclusionTaxonomy}\label{taxonomy}
SBOMs are built around the concept of software components.
While the concept appears intuitive at first glance, a closer look reveals profound ambiguities.
Even well-acknowledged organizations that guide SBOM standardization differ significantly in their interpretations.
For instance, the CRA defines a component as ``software or hardware intended for integration into an electronic information system''~\cite{CRA.2024}, whearas the BSI strictly requires ``a single executable file or an archive file''~\cite{TR-03183-2.2024}.
Meanwhile, the NTIA only considers ``units of software defined by suppliers or authors''~\cite{NTIA.2021e}, and the SPDX refers to ``any unit of content that can be associated with a distribution of software''~\cite{SPDX.package.2024}.
Finally, the EO14028 defines a component as open-source and commercial software components that developers and vendors assemble to create a product~\cite{EO14028}.
Based on these definitions, practical edge cases are assessed differently depending on the definition considered.

For example,
(i) Are APIs considered components? --- According to the EO14028, Yes. According to the BSI, No.
(ii) Are system libraries components? --- According to EO14028, Yes. According to the NTIA, No.
(iii) Is program code that is loaded at runtime a component (sideloading)? --- According to the BSI, Yes. According to the CRA, No.
(iv) Are standard libraries that are bundled with the programming languages a component of the project? --- According to the EO14028, Yes. According to the CRA, No.

If the definition of a component remains ambiguous, SBOM generation tools will inconsistently include or exclude parts of a software project based on the tool developers' subjective interpretations, posing a risk to the supply chain and introducing uncertainty for anyone studying it.
For example, past studies have evaluated entirely different scopes of included elements:
Cofano et al.~\cite{Cofano.2024} measured SBOM quality by equating a component exclusively with a managed dependency, whereas Xiao et al.~\cite{Xiao.2025} defined a component much more widely as any external element contained within a compiled JAR archive. 
Thus, to evaluate the accuracy of SBOM generators, we first need to understand what a component is and what types of components exist.

While there is currently no shared understanding of the term ``component,'' the situation is different when it comes to the overarching goal:
In a recent joint statement, 19 top-level international security agencies defined the value of SBOM in ``securing the software supply chain''~\cite {CISA.2025}.
Thus, to find a definition that can be accepted by as many stakeholders as possible, we take a step back and focus on this overarching goal.

From this perspective, components are elements whose implementation is sourced externally and can introduce vulnerabilities into the resulting software product.
A vulnerability is ``An instance of one or more weaknesses in a Product that can be exploited, causing a negative impact to confidentiality, integrity, or availability; a set of conditions or behaviors that allows the violation of an explicit or implicit security policy''~\cite{MITRE.Glossary}.
This implies that to create such an exploitable weakness in a specific product, an element must be present as executable code in the product.
Assets that are neither shipped with the product nor executable in its runtime environment are not components.

\taxonomy{A component of a software product is any externally sourced part that is shipped as executable code or executed at runtime and thereby being capable of introducing vulnerabilities.}

Our security-focused definition specifies which externally sourced parts of a product we treat as components at an abstract level, but it remains agnostic about the concrete technical mechanisms by which they are introduced.
In the following, we list strategies for incorporating externally sourced code into a software project.
This is, to the best of our knowledge, the first of its kind.

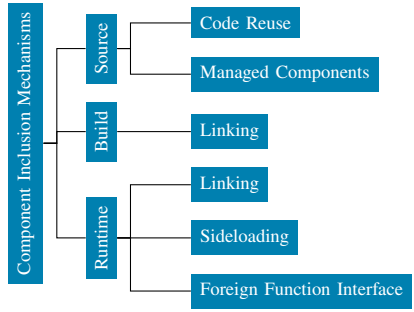
\begin{figure}
    \centering
    \begin{forest}
    forked edges,
    for tree={grow=0,font=\scriptsize,color=white,fill=munsell}
    [Component Inclusion Mechanisms, rotate=90, anchor=south
      [Runtime, rotate=90, anchor=south
        [Foreign Function Interface, anchor=west]
        [Sideloading, anchor=west]
        [Linking, anchor=west]
      ]
      [Build, rotate=90, anchor=south
        [Linking, anchor=west]
      ]
      [Source, rotate=90, anchor=south
        [Managed Components, anchor=west]
        [Code Reuse, anchor=west]
      ]
    ]
    \end{forest}
    \caption{Mechanisms to include software components.}\label{fig:taxonomyTree}
\end{figure}

To follow a structured approach, we will examine, one by one, each stage through which source code can enter a software system and describe possible Component Inclusion Mechanisms (CIM).
The joint approach developed by the BSI and CISA provides a framework for this.
They categorized SBOMs~\cite{CISA.2023b, TR-03183-2.2024} according to the phase of software development in which an SBOM can be generated: design, source, build, analyzed, deployed, or runtime (ref.~\Cref{rel:RoleOfSBOMs}).
In doing so, they list the exact phases in which components can potentially occur.
However, because design precedes any software that may have vulnerabilities, it is out of scope for our definition.
We present an overview of the mechanisms identified in Figure~\ref{fig:taxonomyTree}.

\subsection{Source Components}
The first point of inclusion is the project's source files.
This not only includes the program source code, covering \emph{Code-Reuse}-based components, but also captures any supplementary material, such as project management scripts.

Modern languages provide the developer with tooling to manage a given software project.
These tools depend on a well-defined file format that specifies the direct components and additional build instructions.
As a specification, it is possible to either give a \emph{specific} or \emph{ranged} reference to the required component.
\taxonomy{Source CIMs cover components that can be detected by only analyzing the project source on an arbitrary machine.}

\subsubsection{Code-Reuse}
Code-Reuse components are functionality included in a software project by copying the corresponding source code files or lines.
They are prima facie indistinguishable from the original source code, as they are internal to the software project, even though they originated from an external source.
Those components need not represent the full functionality of the originating component; they may instead comprise cherry-picked elements that the developer used.

A structured approach to code reuse is vendoring, which is natively supported by some languages.
Vendoring occurs when the complete, unmodified source of an external component is embedded into the project's source repository as a distinct unit.
Similar to code reuse, third-party code is incorporated into the project, but unlike code-reuse, the component's structure and origin metadata are preserved, making it identifiable as an external component.

An industry study analyzing 211 million lines of GitHub code suggests that in 2024, 12.3\% of lines of code on GitHub were copy-pasted and therefore constitute code reuse~\cite{Harding.2025}.
Detecting such components is part of an ongoing research effort~\cite{WiSonWWW2022, Kim.2018}, and past reports have shown the negative security impact Code-Reuse can have~\cite{link:codereusevuln, 10.1145/3617555.3617872, Islam.2017, Fischer.2017}.
\taxonomy{Code-Reuse components occur when a developer copies (partial) functionality into their software project.}

\subsubsection{Managed Components}
Most modern languages support package managers such as Gradle for Java~\cite{Gradle.2026}, pip for Python~\cite{PythonPackagingAuthority.2025}, Composer for PHP~\cite{Composer.2026}, or utilize native dependency management integrated directly into the language toolchain, such as Go Mod~\cite{Go.Wiki.2026}.

Most dependency management tools allow not only specifying the component name but also the required version.
Additionally, some package managers allow specifying an exact version (pinned version)~\cite{Gradle.2025b, pip.2025, Poetry.2025, Composer.2025b} but also a version range (constrained version)~\cite{Gradle.2025b, pip.2025, Poetry.2025, Composer.2025b} for greater flexibility.
In some cases, it is also possible not to specify a version at all, which, depending on the package manager, may result in the latest version being loaded (unpinned version)~\cite{Gradle.2025b, pip.2025, Poetry.2025, Composer.2025b}.

Managed dependencies are a widely used approach for integrating components:
In its latest report, Sonatype states that the ecosystems Maven Central, PyPi, npm, and NuGet record around 9.8 trillion downloads annually~\cite{Sonatype.2026}.
E.g., the Python Package Index (PyPi), the index used by pip, provides live statistics and currently hosts around 740,000 projects and records around 120 billion downloads per month~\cite{PyPi.Stats}.
At the same time, examples such as the Shai hulud worm that attacked the npm ecosystem in 2025~\cite{CISA.ShaiHulud.2025}, demonstrated the security implications that can arise when managed dependencies become vulnerable.
\taxonomy{Managed components occur if the source code specifies components using a dedicated package manager.}

\subsection{Build and Analyzed Components}
Build components must be present at build time and before execution, but are not explicitly specified in the source code. 
E.g., if default libraries can be referenced in code, but the precise version used depends on the system's availability.
This covers code that is statically \emph{linked} during building.
While this criterion may also apply to managed components, the crucial discerning factor is that build components depend on the build environment and are not handled in the source code.

CISA and the BSI formally distinguish between Build and Analyzed SBOMs.
While a Build SBOM actively collects component data during compilation, an Analyzed SBOM retrospectively extracts it from finalized artifacts, such as executables, packages, or containers.
However, within the context of our CIMs, this distinction is purely methodological, and no new components are introduced in this stage.
Accordingly, we summarize all components introduced by the compiler as build-component inclusion mechanisms.
\taxonomy{Build CIMs cover any components that depend on the build environment and are present before the first program execution or, if applicable, during compilation, and are not managed.}

\subsubsection{Linking}
Linked components are created when a binary is compiled from source code.
During compilation, the required components are included in the generated binary and thus depend entirely on the build environment.
For example, in Java, the compiler can repackage external libraries into the final application, also known as Uber JAR~\cite{javaUberJar}.
Similarly, in C, external libraries can be copied into the code using the -static flag of gcc~\cite{GNU.OptionsForLinking.2026}.
This approach exists for most compiled programming languages~\cite{Go.2025b,javaUberJar}.
XCodeGhost, in which a trojanized version of Apple's XCode IDE led to a trojanized file being statically linked into applications~\cite{Paloalto.2015}, is a good example of how this mechanism can have security implications.
\taxonomy{Statically linked components are functionalities that are persistently included based on the build environment.}

\subsection{Deployed and Runtime Components}
The deployed SBOM has a similar relationship to the runtime SBOM as the build and analysis SBOM.
While there is a methodological difference when collecting components between statically examining the existing files on a system and analyzing the actual process being executed, it makes no difference for the CIMs.
Therefore, we summarize the components included in the program during execution as Runtime components.

This covers components that are dynamically \emph{linked} during execution, included via \emph{sideloading}, or accessed via \emph{foreign function interfaces}.
The distinguishing feature of dynamic components, relative to managed and especially static components, is that they depend on the runtime environment and on the execution itself.
\taxonomy{Runtime CIMs cover any components that can be added or changed after compilation or during execution, whether manually or automatically, and that are not managed.}

\subsubsection{Linking}
Linked components are those required for execution and loaded from the runtime environment.
They can be shared across different programs and are provided by the runtime environment; for example, \texttt{.so} files on Linux or \texttt{.dll} files on Windows.
However, class paths or JARs specified in a JVM binary invocation also fall into this category. 
This approach is supported by most programming languages~\cite{Oracle.2025, Go.2025c}.

While this form of linking can be useful for avoiding the repeated inclusion of frequently used functionality across the system, it can also have severe security implications.
Particularly complex and serious attacks target this type of component inclusion, e.g., through techniques such as DLL hijacking.
Prominent examples are the SUNBURST attack, in which attackers injected a backdoor into a SolarWinds .dll file~\cite{ncsc.sunburst.2021}, or the Stuxnet attack, in which malicious .dll files were injected into the target systems to compromise the Iranian nuclear program~\cite{CISA.Stuxnet.2013}.
\taxonomy{Runtime-linked components occur when code is dynamically loaded during execution by the runtime environment.}

\subsubsection{Sideloading}
A sideloaded component covers all cases in which code or functionality is dynamically loaded and executed by the program itself.
An example could be a plugin that implements additional functionality on top of a previously defined interface.
Another example would be any language that allows access to and evaluation of external source code during execution.
This approach is supported by some programming languages~\cite{Python.2025b, PHP.eval} and can have security implications.
E.g., in 2020, Autodesk recognized that modified scenes can load a malicious plugin into its 3ds Max application~\cite{Autodesk.2020}.
As a result, malicious actions may occur.
\taxonomy{Sideloaded components occur when code is dynamically loaded during execution by the program.}

\subsubsection{Foreign Function Interface}
Foreign function interfaces allow a programmer to call code written and compiled in a different programming language, enabling access to a wide array of functionality available in the runtime environment.
Most languages ship with such an interface, which can at least handle C-compiled libraries, also known as CFFIs~\cite{javaPanama,phpCFFI}.
This can lead to vulnerabilities being integrated into the program that are uncommon in the main programming language.
For example, the Python Imaging Library (Pillow) uses components written in C that can lead to buffer overflows~\cite{CVE-2021-25290}.
\taxonomy{Foreign Function Interface components occur when the program is calling a function defined in an external library written and compiled in a different programming language.}
\section{Assessment}\label{Benchmark}
We have established the possible CIMs in \Cref{sec:dependencyInclusionTaxonomy}, which provide a high-level overview of the component types.
The question remains which of these are recognized by current SBOM generators.
To address this question, we designed an assessment suite that provides a controlled, reproducible environment in line with our established CIMs.

Our final assessment covers \empirical{\num{225}} individual test executions across \empirical{\num{5}} widely deployed SBOM generators consisting of \empirical{\num{45}} different tests across \empirical{\num{6}} languages (Python, Java, Go, PHP, Rust, C).
We selected the SBOM generators based on the following criteria:
(1) open-source,
(2) general-purpose,
(2) support the selected programming languages,
(3) support the CycloneDX or the SPDX standard,
(4) and be well-known.
Based on these criteria we selected \cdxgen{}, \syft{}, \trivy{}, \ort{}, and \sbomtool{}.
All of them are open source, general-purpose, support the selected programming languages (except \sbomtool{} for PHP), and support CycloneDX or SPDX.
\Syft{} was evaluated by \empirical{\num{15}}, \trivy{} by \empirical{\num{8}}, \cdxgen{} by \empirical{\num{8}}, the \sbomtool{} by \empirical{\num{6}}, and \ort{} by \empirical{\num{5}}  past publications (ref.~\Cref{tab:sbom-tools} in the Appendix).
\Cdxgen{}, \syft{}, \trivy{}, and the \sbomtool{} are also referenced frequently in tooling guidelines~\cite{UpwindSecurity.2025, Sham.2025, Naveen.2025} published by security organizations like the Open-Source Security Foundation.
\ort{} documents its adoption by industry partners such as Deutsche Telekom, HERE, Porsche, Robert Bosch, and Volkswagen on GitHub~\cite{ORT.Adopters.2025}.
An additional reason to include \cdxgen{} in our evaluation was that it is part of OWASP, the publisher of CycloneDX~\cite{CycloneDX}.
To ensure reproducibility, we fixed the tooling versions to \cdxgen{}~12.1.5, \syft{}~v1.42.4, \trivy{}~v0.69.3, \ort{}~83.1.0, and \sbomtool{}~v4.1.5.

We first outline the general considerations of the test suite and subsequently provide details on each test, with the results summarized in Table~\ref{tab:latex_component_detection}.


\definecolor{benchpass}{RGB}{0, 128, 0}    
\definecolor{benchwarn}{RGB}{204, 102, 0} 
\definecolor{benchfail}{RGB}{204, 0, 0}   

\begin{table*}[htbp]
\centering
\scriptsize
\renewcommand{\arraystretch}{1.1}
\begin{tabularx}{1\linewidth}{llYYYYYYYYYY}
\toprule
& & \multicolumn{6}{c}{\textbf{SOURCE}} & \multicolumn{1}{c}{\textbf{BUILD}} & \multicolumn{3}{c}{\textbf{RUNTIME}} \\
\cmidrule(lr){3-8} \cmidrule(lr){9-9} \cmidrule(lr){10-12}
& & \multicolumn{2}{c}{\makecell[b]{Code \\ Reuse}} & \multicolumn{4}{c}{\makecell[b]{Managed \\ Components}} & \multicolumn{1}{c}{\makecell[b]{Linking}} & \multicolumn{1}{c}{\makecell[b]{Cffi}} & \multicolumn{1}{c}{\makecell[b]{Linking}} & \multicolumn{1}{c}{\makecell[b]{Sideloading}} \\
\cmidrule(lr){3-4} \cmidrule(lr){5-8} \cmidrule(lr){9-9} \cmidrule(lr){10-10} \cmidrule(lr){11-11} \cmidrule(lr){12-12}
& & \rotatebox[origin=l]{90}{\makecell[l]{Copy \\ File}\vspace{1mm}} & \rotatebox[origin=l]{90}{\makecell[l]{Vendor \\ Submodule}\vspace{1mm}} & \rotatebox[origin=l]{90}{\makecell[l]{Pinned \\ Version}\vspace{1mm}} & \rotatebox[origin=l]{90}{\makecell[l]{Constrained \\ Version}\vspace{1mm}} & \rotatebox[origin=l]{90}{\makecell[l]{Transitive \\ Shadowing}\vspace{1mm}} & \rotatebox[origin=l]{90}{\makecell[l]{Unpinned \\ Version}\vspace{1mm}} & \rotatebox[origin=l]{90}{Native\vspace{1mm}} & \rotatebox[origin=l]{90}{Libc\vspace{1mm}} & \rotatebox[origin=l]{90}{Native\vspace{1mm}} & \rotatebox[origin=l]{90}{Plugin\vspace{1mm}} \\
\midrule
\multirow{6}{*}{\textbf{Cdxgen}} & Python & \mbox{\Circle} & - & \mbox{\textcolor{benchpass}{\CIRCLE}} & \mbox{\textcolor{benchwarn}{\LEFTcircle}} & \mbox{\textcolor{benchwarn}{\LEFTcircle}} & \mbox{\textcolor{benchwarn}{\LEFTcircle}} & - & \mbox{\Circle} & - & \mbox{\Circle} \\
 & Java & \mbox{\Circle} & - & \mbox{\textcolor{benchpass}{\CIRCLE}} & \mbox{\textcolor{benchwarn}{\LEFTcircle}} & \mbox{\textcolor{benchwarn}{\LEFTcircle}} & \mbox{\textcolor{benchwarn}{\LEFTcircle}} & \mbox{\textcolor{benchpass}{\CIRCLE}} & \mbox{\Circle} & \mbox{\textcolor{benchpass}{\CIRCLE}} & \mbox{\textcolor{benchpass}{\CIRCLE}} \\
 & Go & \mbox{\Circle} & \mbox{\textcolor{benchpass}{\CIRCLE}} & - & \mbox{\textcolor{benchpass}{\CIRCLE}} & - & - & \mbox{\Circle} & \mbox{\Circle} & - & \mbox{\textcolor{benchpass}{\CIRCLE}} \\
 & Php & \mbox{\Circle} & \mbox{\Circle} & \mbox{\textcolor{benchpass}{\CIRCLE}} & \mbox{\textcolor{benchpass}{\CIRCLE}} & \mbox{\textcolor{benchpass}{\CIRCLE}} & \mbox{\textcolor{benchpass}{\CIRCLE}} & - & \mbox{\Circle} & - & \mbox{\Circle} \\
 & Rust & \mbox{\Circle} & - & \mbox{\textcolor{benchwarn}{\LEFTcircle}} & \mbox{\textcolor{benchwarn}{\LEFTcircle}} & \mbox{\textcolor{benchwarn}{\LEFTcircle}} & - & \mbox{\Circle} & \mbox{\Circle} & \mbox{\Circle} & \mbox{\Circle} \\
 & C & \mbox{\Circle} & - & \mbox{\textcolor{benchwarn}{\LEFTcircle}} & \mbox{\textcolor{benchwarn}{\LEFTcircle}} & \mbox{\textcolor{benchwarn}{\LEFTcircle}} & - & \mbox{\Circle} & - & \mbox{\Circle} & \mbox{\Circle} \\
\midrule
\multirow{6}{*}{\textbf{Syft}} & Python & \mbox{\Circle\hspace{1mm}$\mdlgwhtsquare$} & - & \mbox{\textcolor{benchpass}{\CIRCLE}\hspace{1mm}\textcolor{benchpass}{$\mdlgblksquare$}} & \mbox{\Circle\hspace{1mm}$\mdlgwhtsquare$} & \mbox{\Circle\hspace{1mm}$\mdlgwhtsquare$} & \mbox{\Circle\hspace{1mm}$\mdlgwhtsquare$} & - & \mbox{\Circle\hspace{1mm}$\mdlgwhtsquare$} & - & \mbox{\Circle\hspace{1mm}$\mdlgwhtsquare$} \\
 & Java & \mbox{\Circle\hspace{1mm}$\mdlgwhtsquare$} & - & \mbox{\textcolor{benchwarn}{\LEFTcircle}\hspace{1mm}\textcolor{benchwarn}{$\squareleftblack$}} & \mbox{\textcolor{benchwarn}{\LEFTcircle}\hspace{1mm}\textcolor{benchwarn}{$\squareleftblack$}} & \mbox{\textcolor{benchwarn}{\LEFTcircle}\hspace{1mm}\textcolor{benchwarn}{$\squareleftblack$}} & \mbox{\textcolor{benchwarn}{\LEFTcircle}\hspace{1mm}\textcolor{benchwarn}{$\squareleftblack$}} & \mbox{\textcolor{benchwarn}{\LEFTcircle}\hspace{1mm}\textcolor{benchwarn}{$\squareleftblack$}} & \mbox{\Circle\hspace{1mm}$\mdlgwhtsquare$} & \mbox{\textcolor{benchwarn}{\LEFTcircle}\hspace{1mm}\textcolor{benchwarn}{$\squareleftblack$}} & \mbox{\textcolor{benchwarn}{\LEFTcircle}\hspace{1mm}\textcolor{benchwarn}{$\squareleftblack$}} \\
 & Go & \mbox{\Circle\hspace{1mm}$\mdlgwhtsquare$} & \mbox{\textcolor{benchwarn}{\LEFTcircle}\hspace{1mm}\textcolor{benchwarn}{$\squareleftblack$}} & - & \mbox{\textcolor{benchwarn}{\LEFTcircle}\hspace{1mm}\textcolor{benchwarn}{$\squareleftblack$}} & - & - & \mbox{\Circle\hspace{1mm}$\mdlgwhtsquare$} & \mbox{\Circle\hspace{1mm}$\mdlgwhtsquare$} & - & \mbox{\textcolor{benchwarn}{\LEFTcircle}\hspace{1mm}\textcolor{benchwarn}{$\squareleftblack$}} \\
 & Php & \mbox{\Circle\hspace{1mm}$\mdlgwhtsquare$} & \mbox{\Circle\hspace{1mm}$\mdlgwhtsquare$} & \mbox{\textcolor{benchpass}{\CIRCLE}\hspace{1mm}\textcolor{benchpass}{$\mdlgblksquare$}} & \mbox{\textcolor{benchpass}{\CIRCLE}\hspace{1mm}\textcolor{benchpass}{$\mdlgblksquare$}} & \mbox{\textcolor{benchpass}{\CIRCLE}\hspace{1mm}\textcolor{benchpass}{$\mdlgblksquare$}} & \mbox{\textcolor{benchpass}{\CIRCLE}\hspace{1mm}\textcolor{benchpass}{$\mdlgblksquare$}} & - & \mbox{\Circle\hspace{1mm}$\mdlgwhtsquare$} & - & \mbox{\Circle\hspace{1mm}$\mdlgwhtsquare$} \\
 & Rust & \mbox{\Circle\hspace{1mm}$\mdlgwhtsquare$} & - & \mbox{\textcolor{benchwarn}{\LEFTcircle}\hspace{1mm}\textcolor{benchwarn}{$\squareleftblack$}} & \mbox{\textcolor{benchwarn}{\LEFTcircle}\hspace{1mm}\textcolor{benchwarn}{$\squareleftblack$}} & \mbox{\textcolor{benchwarn}{\LEFTcircle}\hspace{1mm}\textcolor{benchwarn}{$\squareleftblack$}} & - & \mbox{\Circle\hspace{1mm}$\mdlgwhtsquare$} & \mbox{\Circle\hspace{1mm}$\mdlgwhtsquare$} & \mbox{\Circle\hspace{1mm}$\mdlgwhtsquare$} & \mbox{\Circle\hspace{1mm}$\mdlgwhtsquare$} \\
 & C & \mbox{\Circle\hspace{1mm}$\mdlgwhtsquare$} & - & \mbox{\textcolor{benchwarn}{\LEFTcircle}\hspace{1mm}\textcolor{benchwarn}{$\squareleftblack$}} & \mbox{\textcolor{benchwarn}{\LEFTcircle}\hspace{1mm}\textcolor{benchwarn}{$\squareleftblack$}} & \mbox{\textcolor{benchpass}{\CIRCLE}\hspace{1mm}\textcolor{benchpass}{$\mdlgblksquare$}} & - & \mbox{\Circle\hspace{1mm}$\mdlgwhtsquare$} & - & \mbox{\Circle\hspace{1mm}$\mdlgwhtsquare$} & \mbox{\Circle\hspace{1mm}$\mdlgwhtsquare$} \\
\midrule
\multirow{6}{*}{\textbf{Trivy}} & Python & \mbox{\Circle\hspace{1mm}$\mdlgwhtsquare$} & - & \mbox{\textcolor{benchwarn}{\LEFTcircle}\hspace{1mm}\textcolor{benchwarn}{$\squareleftblack$}} & \mbox{\Circle\hspace{1mm}$\mdlgwhtsquare$} & \mbox{\Circle\hspace{1mm}$\mdlgwhtsquare$} & \mbox{\Circle\hspace{1mm}$\mdlgwhtsquare$} & - & \mbox{\Circle\hspace{1mm}$\mdlgwhtsquare$} & - & \mbox{\Circle\hspace{1mm}$\mdlgwhtsquare$} \\
 & Java & \mbox{\Circle\hspace{1mm}$\mdlgwhtsquare$} & - & \mbox{\textcolor{benchwarn}{\LEFTcircle}\hspace{1mm}\textcolor{benchwarn}{$\squareleftblack$}} & \mbox{\textcolor{benchwarn}{\LEFTcircle}\hspace{1mm}\textcolor{benchwarn}{$\squareleftblack$}} & \mbox{\textcolor{benchwarn}{\LEFTcircle}\hspace{1mm}\textcolor{benchwarn}{$\squareleftblack$}} & \mbox{\textcolor{benchwarn}{\LEFTcircle}\hspace{1mm}\textcolor{benchwarn}{$\squareleftblack$}} & \mbox{\Circle\hspace{1mm}$\mdlgwhtsquare$} & \mbox{\Circle\hspace{1mm}$\mdlgwhtsquare$} & \mbox{\Circle\hspace{1mm}$\mdlgwhtsquare$} & \mbox{\Circle\hspace{1mm}$\mdlgwhtsquare$} \\
 & Go & \mbox{\Circle\hspace{1mm}$\mdlgwhtsquare$} & \mbox{\textcolor{benchwarn}{\LEFTcircle}\hspace{1mm}\textcolor{benchwarn}{$\squareleftblack$}} & - & \mbox{\textcolor{benchwarn}{\LEFTcircle}\hspace{1mm}\textcolor{benchwarn}{$\squareleftblack$}} & - & - & \mbox{\Circle\hspace{1mm}$\mdlgwhtsquare$} & \mbox{\Circle\hspace{1mm}$\mdlgwhtsquare$} & - & \mbox{\textcolor{benchwarn}{\LEFTcircle}\hspace{1mm}\textcolor{benchwarn}{$\squareleftblack$}} \\
 & Php & \mbox{\Circle\hspace{1mm}$\mdlgwhtsquare$} & \mbox{\Circle\hspace{1mm}$\mdlgwhtsquare$} & \mbox{\textcolor{benchwarn}{\LEFTcircle}\hspace{1mm}\textcolor{benchwarn}{$\squareleftblack$}} & \mbox{\textcolor{benchwarn}{\LEFTcircle}\hspace{1mm}\textcolor{benchwarn}{$\squareleftblack$}} & \mbox{\textcolor{benchwarn}{\LEFTcircle}\hspace{1mm}\textcolor{benchwarn}{$\squareleftblack$}} & \mbox{\textcolor{benchwarn}{\LEFTcircle}\hspace{1mm}\textcolor{benchwarn}{$\squareleftblack$}} & - & \mbox{\Circle\hspace{1mm}$\mdlgwhtsquare$} & - & \mbox{\Circle\hspace{1mm}$\mdlgwhtsquare$} \\
 & Rust & \mbox{\Circle\hspace{1mm}$\mdlgwhtsquare$} & - & \mbox{\textcolor{benchwarn}{\LEFTcircle}\hspace{1mm}\textcolor{benchwarn}{$\squareleftblack$}} & \mbox{\textcolor{benchwarn}{\LEFTcircle}\hspace{1mm}\textcolor{benchwarn}{$\squareleftblack$}} & \mbox{\textcolor{benchwarn}{\LEFTcircle}\hspace{1mm}\textcolor{benchwarn}{$\squareleftblack$}} & - & \mbox{\Circle\hspace{1mm}$\mdlgwhtsquare$} & \mbox{\Circle\hspace{1mm}$\mdlgwhtsquare$} & \mbox{\Circle\hspace{1mm}$\mdlgwhtsquare$} & \mbox{\Circle\hspace{1mm}$\mdlgwhtsquare$} \\
 & C & \mbox{\Circle\hspace{1mm}$\mdlgwhtsquare$} & - & \mbox{\textcolor{benchwarn}{\LEFTcircle}\hspace{1mm}\textcolor{benchwarn}{$\squareleftblack$}} & \mbox{\textcolor{benchwarn}{\LEFTcircle}\hspace{1mm}\textcolor{benchwarn}{$\squareleftblack$}} & \mbox{\textcolor{benchwarn}{\LEFTcircle}\hspace{1mm}\textcolor{benchwarn}{$\squareleftblack$}} & - & \mbox{\Circle\hspace{1mm}$\mdlgwhtsquare$} & - & \mbox{\Circle\hspace{1mm}$\mdlgwhtsquare$} & \mbox{\Circle\hspace{1mm}$\mdlgwhtsquare$} \\
\midrule
\multirow{6}{*}{\textbf{Ort}} & Python & \mbox{\Circle\hspace{1mm}$\mdlgwhtsquare$} & - & \mbox{\textcolor{benchpass}{\CIRCLE}\hspace{1mm}\textcolor{benchwarn}{$\squareleftblack$}} & \mbox{\textcolor{benchpass}{\CIRCLE}\hspace{1mm}\textcolor{benchwarn}{$\squareleftblack$}} & \mbox{\textcolor{benchpass}{\CIRCLE}\hspace{1mm}\textcolor{benchwarn}{$\squareleftblack$}} & \mbox{\textcolor{benchpass}{\CIRCLE}\hspace{1mm}\textcolor{benchwarn}{$\squareleftblack$}} & - & \mbox{\Circle\hspace{1mm}$\mdlgwhtsquare$} & - & \mbox{\Circle\hspace{1mm}$\mdlgwhtsquare$} \\
 & Java & \mbox{\Circle\hspace{1mm}$\mdlgwhtsquare$} & - & \mbox{\textcolor{benchpass}{\CIRCLE}\hspace{1mm}\textcolor{benchwarn}{$\squareleftblack$}} & \mbox{\textcolor{benchpass}{\CIRCLE}\hspace{1mm}\textcolor{benchwarn}{$\squareleftblack$}} & \mbox{\textcolor{benchpass}{\CIRCLE}\hspace{1mm}\textcolor{benchwarn}{$\squareleftblack$}} & \mbox{\textcolor{benchpass}{\CIRCLE}\hspace{1mm}\textcolor{benchwarn}{$\squareleftblack$}} & \mbox{\Circle\hspace{1mm}$\mdlgwhtsquare$} & \mbox{\Circle\hspace{1mm}$\mdlgwhtsquare$} & \mbox{\Circle\hspace{1mm}$\mdlgwhtsquare$} & \mbox{\Circle\hspace{1mm}$\mdlgwhtsquare$} \\
 & Go & \mbox{\Circle\hspace{1mm}$\mdlgwhtsquare$} & \mbox{\textcolor{benchpass}{\CIRCLE}\hspace{1mm}\textcolor{benchwarn}{$\squareleftblack$}} & - & \mbox{\textcolor{benchpass}{\CIRCLE}\hspace{1mm}\textcolor{benchwarn}{$\squareleftblack$}} & - & - & \mbox{\Circle\hspace{1mm}$\mdlgwhtsquare$} & \mbox{\Circle\hspace{1mm}$\mdlgwhtsquare$} & - & \mbox{\textcolor{benchpass}{\CIRCLE}\hspace{1mm}\textcolor{benchwarn}{$\squareleftblack$}} \\
 & Php & \mbox{\Circle\hspace{1mm}$\mdlgwhtsquare$} & \mbox{\textcolor{benchfail}{E}\hspace{1mm}\textcolor{benchfail}{E}} & \mbox{\textcolor{benchpass}{\CIRCLE}\hspace{1mm}\textcolor{benchwarn}{$\squareleftblack$}} & \mbox{\textcolor{benchpass}{\CIRCLE}\hspace{1mm}\textcolor{benchwarn}{$\squareleftblack$}} & \mbox{\textcolor{benchpass}{\CIRCLE}\hspace{1mm}\textcolor{benchwarn}{$\squareleftblack$}} & \mbox{\textcolor{benchpass}{\CIRCLE}\hspace{1mm}\textcolor{benchwarn}{$\squareleftblack$}} & - & \mbox{\Circle\hspace{1mm}$\mdlgwhtsquare$} & - & \mbox{\Circle\hspace{1mm}$\mdlgwhtsquare$} \\
 & Rust & \mbox{\Circle\hspace{1mm}$\mdlgwhtsquare$} & - & \mbox{\textcolor{benchpass}{\CIRCLE}\hspace{1mm}\textcolor{benchwarn}{$\squareleftblack$}} & \mbox{\textcolor{benchpass}{\CIRCLE}\hspace{1mm}\textcolor{benchwarn}{$\squareleftblack$}} & \mbox{\textcolor{benchpass}{\CIRCLE}\hspace{1mm}\textcolor{benchwarn}{$\squareleftblack$}} & - & \mbox{\Circle\hspace{1mm}$\mdlgwhtsquare$} & \mbox{\Circle\hspace{1mm}$\mdlgwhtsquare$} & \mbox{\Circle\hspace{1mm}$\mdlgwhtsquare$} & \mbox{\Circle\hspace{1mm}$\mdlgwhtsquare$} \\
 & C & \mbox{\Circle\hspace{1mm}$\mdlgwhtsquare$} & - & \mbox{\textcolor{benchfail}{E}\hspace{1mm}\textcolor{benchfail}{E}} & \mbox{\textcolor{benchfail}{E}\hspace{1mm}\textcolor{benchfail}{E}} & \mbox{\textcolor{benchfail}{E}\hspace{1mm}\textcolor{benchfail}{E}} & - & \mbox{\Circle\hspace{1mm}$\mdlgwhtsquare$} & - & \mbox{\Circle\hspace{1mm}$\mdlgwhtsquare$} & \mbox{\Circle\hspace{1mm}$\mdlgwhtsquare$} \\
\midrule
\multirow{6}{*}{\textbf{sbom tool}} & Python & \mbox{$\mdlgwhtsquare$} & - & \mbox{\textcolor{benchwarn}{$\squareleftblack$}} & \mbox{\textcolor{benchwarn}{$\squareleftblack$}} & \mbox{\textcolor{benchwarn}{$\squareleftblack$}} & \mbox{\textcolor{benchwarn}{$\squareleftblack$}} & - & \mbox{$\mdlgwhtsquare$} & - & \mbox{$\mdlgwhtsquare$} \\
 & Java & \mbox{$\mdlgwhtsquare$} & - & \mbox{\textcolor{benchwarn}{$\squareleftblack$}} & \mbox{\textcolor{benchwarn}{$\squareleftblack$}} & \mbox{\textcolor{benchwarn}{$\squareleftblack$}} & \mbox{\textcolor{benchwarn}{$\squareleftblack$}} & \mbox{$\mdlgwhtsquare$} & \mbox{$\mdlgwhtsquare$} & \mbox{$\mdlgwhtsquare$} & \mbox{$\mdlgwhtsquare$} \\
 & Go & \mbox{$\mdlgwhtsquare$} & \mbox{\textcolor{benchwarn}{$\squareleftblack$}} & - & \mbox{\textcolor{benchwarn}{$\squareleftblack$}} & - & - & \mbox{$\mdlgwhtsquare$} & \mbox{$\mdlgwhtsquare$} & - & \mbox{\textcolor{benchwarn}{$\squareleftblack$}} \\
 & Php & \mbox{$\mdlgwhtsquare$} & \mbox{$\mdlgwhtsquare$} & \mbox{$\mdlgwhtsquare$} & \mbox{$\mdlgwhtsquare$} & \mbox{$\mdlgwhtsquare$} & \mbox{$\mdlgwhtsquare$} & - & \mbox{$\mdlgwhtsquare$} & - & \mbox{$\mdlgwhtsquare$} \\
 & Rust & \mbox{$\mdlgwhtsquare$} & - & \mbox{\textcolor{benchwarn}{$\squareleftblack$}} & \mbox{\textcolor{benchwarn}{$\squareleftblack$}} & \mbox{\textcolor{benchwarn}{$\squareleftblack$}} & - & \mbox{$\mdlgwhtsquare$} & \mbox{$\mdlgwhtsquare$} & \mbox{$\mdlgwhtsquare$} & \mbox{$\mdlgwhtsquare$} \\
 & C & \mbox{$\mdlgwhtsquare$} & - & \mbox{$\mdlgwhtsquare$} & \mbox{$\mdlgwhtsquare$} & \mbox{$\mdlgwhtsquare$} & - & \mbox{$\mdlgwhtsquare$} & - & \mbox{$\mdlgwhtsquare$} & \mbox{$\mdlgwhtsquare$} \\
\bottomrule
\end{tabularx}
\caption{Aggregated Component Detection Accuracy. Circles ($\mdlgblkcircle$) represent CycloneDX, Squares ($\mdlgblksquare$) represent SPDX. Colors: \textcolor{benchpass}{Green} (Exact Match), \textcolor{benchwarn}{Orange} (+FP), \textcolor{benchfail}{Red} (Partial), outline (None/Empty), and \textcolor{benchfail}{E} (Error/Crash).}
\label{tab:latex_component_detection}
\end{table*}

\subsection{General Considerations}
To ensure full control and knowledge of any introduced component, we created our own packages for each language and made them available to the management tools of the languages we tested.
We programmed two custom packages \texttt{Dependency-One} and \texttt{Dependency-Two}.
\texttt{Dependency-One} is available in multiple versions, whereas \texttt{Dependency-Two} package depends on a fixed version of \texttt{Dependency-One}.
The packages solely provide the functionality to output their name and version, and the tests use this functionality to generate output.
Thus, we can derive a ground truth of the components used to build and run each test project.
Our custom packages do not rely on any additional third-party dependencies and should consequently be the only components appearing in any generated SBOM.

\subsubsection{The naming problem}
\label{sec:naming_problem}
A well-known challenge in handling SBOMs is the ``naming problem'', where component names lack unambiguity and cross-ecosystem standardization~\cite{NTIA.Naming.2021, Meyers.2023, Brabandere.2025, Alrich.2020}.
For example, in our tests, different tools specified the name of our dependency as ``io.github.sbenchmark.java-dependency-one'', ``io.github.sbenchmark:java-dependency-one'', or ``java-dependency-one'' for the same Java test.
As a result, simple string comparisons may fail.
Instead of a simple comparison, we use the name field's content and define various spellings as aliases in our expected ground truth.
We manually checked all SBOMs that deviated from our expected ground truth, verified the correct implementation, and added additional aliases where necessary.

\subsubsection{Ground Truth Generation}
The versions of the components used may depend on the system (e.g., CFFI test) or the behavior of the package manager (e.g., unpinned version test).
To ensure that the ground truth remains accurate at all times, we update it dynamically within the Docker container - the same environment in which the SBOM generators are executed.
To do this, the test applications are built and executed before the tests.
The printed names and versions of the components are parsed and stored in the ground truth file.
Afterward, all build artifacts are deleted.

\subsubsection{Component Types}
Although CycloneDX is often referred to as an SBOM standard, it provides ``xBOM Capabilities'', representing different types of Bill of Materials (BOM), such as Software BOMs (SBOM), Cryptography BOMs (CBOM), or Hardware BOMs (HBOM)~\cite{ECMA.CycloneDX.2025}.
This is achieved, among other things, by having a component have different types, such as ``application'', ``container'', ``file'', or ``cryptographic asset''~\cite{ECMA.CycloneDX.2025}.
This paper focuses on SBOMs, so the assessment only considers the types ``application'', ``framework'', and ``library''.
Otherwise, components of other types that may also be detected but do not appear in our ground truth would be identified as false positives.

\subsubsection{SBOM Processing}
Our tool uses the official Python libraries for CycloneDX~\cite{CycloneDX.PythonLibrary.2026} and SPDX~\cite{SPDX.Python.Tools.2026} to parse and validate the SBOMs.
Therefore, the tool's support is limited to the standard versions supported by the libraries as of the date of testing: CycloneDX 1.0 to 1.7~\cite{CycloneDX.PythonLibrary.Docs.2026} and SPDX 2.2 to 2.3~\cite{SPDX.Python.Tools.Documentation.2026}. 
For both standards, we provide JSON and XML support.

\subsubsection{Files provided}
We found that some tools rely heavily on lock files (e.g., pipfile.lock, gradle.lock, Gopkg.lock, or composer.lock) to function.
Further, we found conflicting recommendations regarding whether they should be included in code repositories~\cite{GitHub.2010, Composer.2025c, Atlassian.2025}.
To give the tools the best chance of recognizing components, we decided in favor of the tools and generated these files.
Note that they may not always be available in practice, which can lead to a lower detection rate.
For the same reason, we also created the go.sum files, as they are recommended for pushing to repositories, even if they are not generated automatically~\cite{Go.Wiki.2026}.
However, we will discuss the implications of the availability of such files in detail in Section~\ref{dis:Accuracy}.

\subsection{General Findings}
In addition to findings from specific tests, we have identified several general patterns in the tools’ behavior worth noting.
Some tools always list certain files or projects as components of the type ``Application'', ``library'', or ``Unknown'', which we would not classify as such and therefore count as false positives.
For example, \trivy{} listed lock files like the \texttt{Cargo.lock} or the \texttt{go.mod} as a component of the type ``Application'' or \ort{} the \texttt{composer.json}.
For go tests, \trivy{} and \syft{} identified the package name specified in the \texttt{go.mod} as a component (e.g. ``example.com/cgo-libc'').
For Java tests, \syft{} listed the \texttt{gradle-wrapper} as a component.
Additionally, all tools except the \sbomtool{} listed Rust itself as a component of the project for at least one SBOM standard.

All SPDX SBOMs include a component that describes the project itself.
However, the components of \syft{} and \trivy{} are filtered by our criteria, which require that we consider only components of type application, library, or unknown if a specification was available.
They used the SPDX-2.3 format by default instead of SPDX-2.2, which allowed the field \texttt{primaryPackagePurpose}, which was set to \texttt{FILE}, respectively \texttt{SOURCE}.
\Ort{} and the \sbomtool{} used SPDX-2.2 by default, which does not allow the field, leading to false-positives.

Additionally, we found configuration issues: even though it was specified as optional on the help pages, \sbomtool{} required additional options, such as the \texttt{package name}, or it would fail with an error for every test.
\Ort{} failed with an error for some tests, when a lock file was expected but not identified.

\subsection{Source Tests}
This category contains all tests that cover source-based Component Inclusion Mechanism~(ref.~\Cref{fig:taxonomyTree}).
The Code-Reuse test is straightforward and offers two implementations.
For the direct test (\emph{Copy-File}), we take the source code of \texttt{Dependency-One} and copy its source file into the source tree of the test project.
Such direct inclusion of source code reflects a developer copying and pasting the desired functionality into their code.
The second test only works if the language supports a dependency folder where the raw source code for project dependencies is placed, e.g., by the build tool.
If such a folder exists, we copy our custom components into it, but do not add an entry to any management file.
This can, for example, be observed in PHP projects that do contain a vendor folder in their source tree.
We named this test \emph{Vendor-Submodule}.

The second type of tests covers managed components.
\texttt{Dependency-One} was installed using a package manager:
(a) by specifying an exact version (pinned version),
(b) by specifying a version range (constrained version),
and (c) without specifying a version.
In addition, components may depend on one another and require specific versions. We tested this scenario by allowing different versions of Dependency-One in the package manager’s specification.
However, we also include Dependency-Two, which requires a specific version of Dependency-One.
Therefore, parsing the package manager manifest alone is not sufficient to identify the correct version number.

We selected one popular package manager for each language.
For Python, we selected \texttt{pip}~\cite{PythonPackagingAuthority.2025}, because all our SBOM generators support it, and its corresponding \texttt{requirements.txt} file is widely used.
The format of the \texttt{requirements.txt} allows for both, providing only the name of the required package, as well as giving ranges~\cite{pip.requirements.txt}.
Regarding Java, we chose Gradle~\cite{Gradle.2026}, because it also met all our requirements.
According to the JetBrains Developer Ecosystem Survey 2023~\cite{JetBrains.2023}, \empirical{\num{46}\%} of Java developers use Gradle~\cite{Gradle.2026} regularly.
Gradle defines the build process in a \texttt{gradle.build}~\cite{Gradle.build.process}.
Go exclusively manages dependencies via the \texttt{go.mod} file~\cite{Go.mod.reference}.
The \texttt{go.mod} file specifies the required packages by providing a path and the minimum required version.
Go then determines the actual version used via minimal version selection~\cite{Go.mvs}.
In broad terms, Go resolves all components and their transitive dependencies, and then selects the lowest version that fulfills the requirements of all included packages.
Thus, if a developer provides only a single component, the reference is specific, but as soon as the specification includes multiple components with shared components across different versions, the reference becomes ranged.
Rust exhibits near-identical behavior and exclusively uses cargo and a corresponding \texttt{Cargo.toml} file to specify dependencies~\cite{Rust.manifest.2026}.
Consequently, we have tests for specific references and for ranged references, in which we observe what happens when a second component requires a different but compatible version of a directly included component.
To manage dependencies in PHP, we use Composer~\cite{Composer.2026}, the dependency management tool referenced in the official PHP documentation~\cite{PHP.Composer}.
Composer relies on a \texttt{composer.json} configuration file supporting our requirements for specific and ranged tests.
For C, we selected Conan~\cite{Conan.2026} as the package manager, as it is the package manager claimed to be supported by all SBOM generators (See Table~\ref{tab:tool_capabilities}).
It utilizes a \texttt{conanfile.py} or \texttt{conanfile.txt} and requires that either exact versions or explicit version ranges are specified.
The highest available version that satisfied all active constraints is used~\cite{Conan2.versionranges}.

\textbf{Results.}
No tool passed the Copy-File test.
All tools passed the Vendor-Submodule test for Go, but not for PHP.
However, \ort{} fails with an error because the \texttt{composer.lock} file is not available in the vendor-submodule folder.

Most tests for managed dependencies were passed by all SBOM generators.
\Ort{} failed with an error during the C tests, as it did not recognize the Conan lockfile.
The lockfile name must be configured in the corresponding option in \ort{}.
However, in many cases, false positives were detected alongside our expected components.
For example, \trivy{} listed the \texttt{.lock} files as ``application'' components, and \cdxgen{}, \ort{}, \syft{}, and \trivy{} listed Rust itself as a component.
\Cdxgen{} listed additional wrong versions for Python and Java for the constrained version, transitive shadowing, and unpinned tests.
This included the exact version notion specified within the manifests, e.g., version \texttt{+}, or version \texttt{[0.0.0}.
\Syft{} and \trivy{} failed in all Python tests except the pinned version, as their default behavior is to not guess unspecified versions~\cite{Trivy.Languages, Anchore.Docs.2026}.

\subsection{Build Tests}
This category contains all tests for Build Component Inclusion Methods~(ref.~\Cref{fig:taxonomyTree}).
To ensure complete control and knowledge of any introduced component, we again rely on our custom components.
If possible, we reuse the packages or binaries already created for managed components, but do not leverage any central dependency repository for access and statically link \texttt{Dependency-One} into the program during the build process.
As Python and PHP are interpreted languages, no linking occurs, and the test is skipped.
Go dropped support for binary packages in 2019~\cite{link:github:goNoBinaryPackage, link:googlesources:goNoBinaryPackage}.
Consequently, we could not write a dedicated test that creates a binary package for our component and then statically links against it.
Rust, however, does support creating libraries that can be statically linked.
The compiler specification, including the correct binary to link, is provided in a \texttt{build.rs} file.
The closest Java and JVM programs come to it is including all precompiled dependencies in the 'uber jar'.
We thus added a system scope component for \texttt{Dependency-One} to our build file, which was then included in the generated JAR.
C supports static linking, e.g., by integrating header files.

\textbf{Results.} 
For Java, \cdxgen{} and \syft{} identify the component as expected. 
However, \syft{} additionally lists the gradle-wrapper as a component. 
In all other cases, our expected components were not detected.

\subsection{Runtime Tests}
This category contains all tests for the Runtime Inclusion Methods~(ref.~\Cref{fig:taxonomyTree}).
We leverage libc for our CFFI test.
Libc is present on every system and has no further dependencies.
We call the function \texttt{gnu\_get\_libc\_version} in our tests.
Every language except C itself supports CFFI and thus has a corresponding test.
PHP additionally needs to activate the corresponding module via a flag in its init file.
Given that we have not observed any SBOM generator accessing the ini file, this should not affect detection capabilities.

Linking during runtime is not available for Python and PHP, as those are interpreted languages that do not produce binaries.
Additionally, Go dropped support for binary packages in 2019~\cite{link:github:goNoBinaryPackage,link:googlesources:goNoBinaryPackage}.
We included \texttt{Dependency-One} in the program at runtime via a Java classpath flag, specified the linking instructions in the \texttt{build.rs} file for Rust, and created a dynamic library in C.

For Sideloading, we dynamically loaded the functionality of \texttt{Dependency-One} during runtime.
Such runtime loading may involve importing code from the Internet or loading binary code.
The test is available for all programming languages.
For Python, we use the eval functionality to inject a plugin. 
In Java, we dynamically loaded the contents of \texttt{Dependency-One} and executed its main functionality.
The path to the corresponding jar is hard-coded.
In PHP, the feature that loads components at runtime via a URL as an include path is disabled by default.
However, it can be activated by setting the corresponding parameter in the \texttt{php.ini}, and we implemented our test using the \texttt{include} command with a URL pointing to the raw data file for \texttt{Dependency-One}.
Go provides built-in functionality for loading plugins, which we leveraged, and Rust simply allows dynamically loading dynamic library binaries at runtime using \texttt{dlopen} and \texttt{dlsym}.
Similar to Rust, we utilized the \texttt{dlopen} functionality for C.

\textbf{Results.}
No tool recognized the use of libc and therefore failed the CFFI test.
Both \cdxgen{} and \syft{} detected the linked and sideloaded component, which was included in a separate plugin subfolder within the test folder.
\Syft{} additionally listed the gradle-wrapper for Java as a component and a component called ``command-line-arguments'' for Go.
All tools detected the Go plugin, even though some also listed false positives.
However, only \cdxgen{} and \syft{} detected the Java plugin.

\section{Manual SBOM Generator Analysis}
\label{sec:generatorBisection}

\begin{table*}[ht]
    \centering
    \scriptsize
    \begin{threeparttable}
        \newcommand{\faCloseGray}{\textcolor{gray}{\faClose}}

\begin{tabularx}{\textwidth}{l|XX:X:XXXXXX|X:XXXXX|XX|XX|XX|XXXXXX}
        & \multicolumn{9}{c|}{\textbf{Python}} & \multicolumn{6}{c|}{\textbf{Java}} & \multicolumn{2}{c|}{\textbf{Go}} & \multicolumn{2}{c|}{\textbf{PHP}} & \multicolumn{2}{c|}{\textbf{Rust}} & \multicolumn{6}{c}{\textbf{C}} \\
        \midrule
        
            & \rotatebox{90}{.py} & \rotatebox{90}{wheel/egg} & \rotatebox{90}{Pip} & \rotatebox{90}{Poetry} & \rotatebox{90}{pdm} & \rotatebox{90}{Hatch} & \rotatebox{90}{uv} & \rotatebox{90}{Pixi} & \rotatebox{90}{Conda} & \rotatebox{90}{JVM archives~} & \rotatebox{90}{Maven} & \rotatebox{90}{Gradle} & \rotatebox{90}{Quarkus} & \rotatebox{90}{Mill} & \rotatebox{90}{SBT} & \rotatebox{90}{Go modules} & \rotatebox{90}{Go binaries} & \rotatebox{90}{Composer} & \rotatebox{90}{Pear/Pecl} & \rotatebox{90}{Cargo} & \rotatebox{90}{Rust binaries} & \rotatebox{90}{Conan} & \rotatebox{90}{CMake} & \rotatebox{90}{Meson} & \rotatebox{90}{vcpkg} & \rotatebox{90}{C Source} & \rotatebox{90}{C binaries}\\
    \midrule
    \textbf{Cdxgen} 
            & \faCheck{} & \faCheck{} & \faCheck{} & \faCheck{} & \faCheck{} & \faCheck{} & \faCheck{} & \faCheck{} & \faCheck{} 
            & \faCheck{} & \faCheck{} & \faCheck{} & \faCheck{} & \faCheck{} & \faCheck{}
            & \faCheck{} & \faCheck{} 
            & \faCheck{} & \faCloseGray{}
            & \faCheck{} & \faCheck{} 
            & \faCheck{} & \faCheck{} & \faCheck{} & \faCheck{} & \faCheck{} & \faCheck{}\\
    \textbf{Syft} 
        & \faCloseGray{} & \faCheck{} & \faCheck{} & \faCheck{} & \faCheck{} & \faCloseGray{} & \faCheck{} & \faCloseGray{} & \faCheck{} 
        & \faCheck{} & \faCloseGray{} & \faCheck{} & \faCloseGray{} & \faCloseGray{} & \faCloseGray{}
        & \faCheck{} & \faCheck{} 
        & \faCheck{} & \faCheck{}
        & \faCheck{} & \faCheck{} 
        & \faCheck{} & \faCloseGray{} & \faCloseGray{} & \faCloseGray{} & \faCloseGray{} & \faCheck{}\\
    \textbf{Trivy}
        & \faCloseGray{} & \faCheck{} & \faCheck{} & \faCheck{} & \faCloseGray{} & \faCloseGray{} & \faCheck{} & \faCloseGray{} & \faCheck{} 
        & \faCheck{} & \faCheck{} & \faCheck{} & \faCloseGray{} & \faCloseGray{} & \faCheck{}
        & \faCheck{} & \faCheck{} 
        & \faCheck{} & \faCloseGray{}
        & \faCheck{} & \faCheck{} 
        & \faCheck{} & \faCloseGray{} & \faCloseGray{} & \faCloseGray{} & \faCloseGray{} & \faCloseGray{}\\
    \textbf{ORT} 
        & \faCloseGray{} & \faCloseGray{} & \faCheck{} & \faCheck{} & \faCloseGray{} & \faCloseGray{} & \faCloseGray{} & \faCloseGray{} & \faCheck{} 
        & \faCloseGray{} & \faCheck{} & \faCheck{} & \faCloseGray{} & \faCloseGray{} & \faCheck{}
        & \faCheck{} & \faCloseGray{} 
        & \faCheck{} & \faCloseGray{}
        & \faCheck{} & \faCloseGray{} 
        & \faCheck{} & \faCloseGray{} & \faCloseGray{} & \faCloseGray{} & \faCloseGray{} & \faCloseGray{}\\
    \textbf{SBOM Tool} 
        & \faCloseGray{} & \faCloseGray{} & \faCheck{} & \faCheck{} & \faCloseGray{} & \faCloseGray{} & \faCloseGray{} & \faCloseGray{} & \faCheck{} 
        & \faCloseGray{} & \faCheck{} & \faCheck{} & \faCloseGray{} & \faCloseGray{} & \faCloseGray{}
        & \faCheck{} & \faCloseGray{} 
        & \faCloseGray{} & \faCloseGray{}
        & \faCheck{} & \faCloseGray{} 
        & \faCheck{} & \faCloseGray{} & \faCloseGray{} & \faCheck{} & \faCloseGray{} & \faCloseGray{}\\
\end{tabularx}
        \caption{High-Level Overview of the approaches that the tools use to identify components (grouped).}
        \label{tab:tool_capabilities}
    \end{threeparttable}
\end{table*}

So far, we have identified component inclusion mechanisms and evaluated whether SBOM generators can detect components integrated through these mechanisms.
While our assessment automatically identifies \emph{what} shortcomings exist, it does not initially explain \emph{why} they occur.
However, it is important to find a rationale for the results.
Therefore, we conducted a manual code analysis of the tools under investigation to determine how they identify components.
Due to space constraints, we only cover the high-level processes in this section.
We provide a high-level overview of the capabilities based on the documentation in Table~\ref{tab:tool_capabilities} and more details on the specific files and data used by the tools in \Cref{tbl:analyzedFiles} in the Appendix.

We will lead our bisection of the SBOM generators with our general insights (\Cref{sec:generatorBistection:general}) and subsequently detail specific insights unique to \cdxgen{} (\Cref{sec:generatorBistection:cdxgen}), \syft{} (\Cref{sec:generatorBistection:syft}), \trivy{} (\Cref{sec:generatorBistection:trivy}), \ort{} (\Cref{sec:generatorBistection:ort}), and the \sbomtool{} (\Cref{sec:generatorBistection:sbomtool}).

\subsection{General Observations}\label{sec:generatorBistection:general}
Below, we describe the key findings of the source code analysis.
We give a detailed listing of what analysis approaches are supported by which tool in \Cref{tbl:analyzedFiles}.

\subsubsection{High Support for Package Management}
SBOM generators rely heavily on project managers to detect components.
To do this, they focus on manifest files or lock files.
These files often have a standardized structure, making them easy to parse.
Some tools also query the project management tools themselves or dynamically generate the relevant files by running the projects.

\subsubsection{Intermediate Support for Build Artifacts}
In addition to parsing \cdxgen{}, \syft{}, and \trivy{} analyzed binaries and archives, such as Java jar files or Python egg files.
This makes it possible not only to generate an SBOM based on source code files - what CISA and the BSI refer to as a ``Source SBOM " ~\cite{CISA.2023b, TR-03183-2.2024} - but also based on projects that have already been built.
This means that, at least technically, it is possible to generate SBOMs of the ``Build'' or ``Analyze'' type.

\subsubsection{Low Support for Source Analysis}
Only \cdxgen{} actually scanned the source code files for dependencies, and only when non-default flags were selected.
In contrast, \cdxgen{}, \syft{}, \trivy{}, and \ort{} were capable of running the build script or package managers to identify components not specified in manifests or lock files.
However, in most cases, this was only done after selecting specific options.

\subsection{\Cdxgen}\label{sec:generatorBistection:cdxgen}
CycloneDX Generator (\cdxgen{})~\cite{cdxgen} supported the most analysis approaches across all examined SBOM generators.
For Python, \cdxgen{} primarily supports parsing lockfiles and manifest files
and enriched them with data available on PyPi.
In addition to its official documentation, it also supported Pixi~\cite[line: 3681 ff.]{cdxgen.index.js}.
If the install-deps flag is enabled, it can create a venv, install the packages specified in the manifest files, and run pip freeze to identify components or query package managers like conda to gather information.
If the tool is executed in an existing venv, that venv is used.
Further, a deep mode can be enabled, which offers additional capabilities such as inspecting the actual source code.
Build components specified in PYTHON\_EXCLUDED\_COMPONENTS~\cite[line: 384 ff.]{cdxgen.utils.js} are ignored.
For Java, components were identified through static analysis of JVM archives and project manifests, supplemented by lock files and build tool queries to Maven, Gradle, or SBT.
In Go, \cdxgen{} first checked whether the provided file path points to a binary.
If this were the case, BLint~\cite{OWASP.Blint}, a tool from OWASP designed to evaluate executables, was applied, and its output was used.
BLint parsed archives directly and utilized the LIEF library for binaries~\cite{Thomas.2025}.
Otherwise, the Go CLI is used to extract components from the go.mod file.
Additionally, project management files are parsed if they exist.
For PHP, high-level metadata was extracted from the \texttt{composer.json} file, but the components are identified only on composer.lock files.
\Cdxgen{} supports generating missing lock files when the corresponding option is enabled.
For Rust, cdxgen parsed the corresponding project management files. 
If dependency installation was enabled, it also used cargo commands to generate lock files. 
As with Go, Rust binary files can be analyzed using Blint.
For C, Conan package manager files, CMake files, Meson files, and VCpkg files are parsed.
Further, if the deep mode is activated, \cdxgen{} is capable of utilizing AppThreat Atom~\cite{Atom.2026}, a static analysis tool that can scan \texttt{.c} and \texttt{.h} files to detect include statements.
Utilizing osquery~\cite{Osquery.2026}, it identifies installed binaries and headers on the operating system level, to match them against the include statements.
Finally, C binaries could be scanned using Blint.

Notably, while manually analyzing the source code, we came across a developer comment reading ``The position taken by cdxgen is `Some SBOM is better than no SBOM'''~\cite[line: 15739]{cdxgen.utils.js}, providing insight into the design philosophies of \cdxgen{}.

\subsection{\Syft}\label{sec:generatorBistection:syft}
For Python, \syft{} processed installation directories, manifest files, and package manager-specific files.
However, only components with pinned versions are listed by default.
Version guessing is available if the option is set.
Additionally, it searches for \texttt{pyvenv.cfg} files, which are part of venvs, to extract metadata, like the Python version, and searches the \texttt{conda-meta} folder for JSON files.
For Java, the metadata within archives is parsed as well as wrapped archives (tar-based and zip-based).
Further, \texttt{pom.xml} files, which are used, e.g., by Maven, are supported as is \texttt{gradle.lockfile}.
Finally, runtime versions were extracted from JRE and JVM environments, and components compiled into GraalVM executables could be extracted.
For Go, PHP, C, and Rust the project management files are used with the addition of the binary analysis for C, Go, and Rust.

\subsection{\Trivy}\label{sec:generatorBistection:trivy}
For Python, \trivy{} supported parsing \texttt{requirements.txt} files, but only components with pinned versions were listed, unless a special flag was set.
Additionally, other common manifest files and metadata within installed packages are parsed.
For Java, \trivy{} supported the analysis of compiled archives by extracting embedded metadata, as well as the parsing of manifest files.
For PHP and C, \trivy{} relies solely on parsing the project management files. 
For Go and Rust, both management files and binary analysis are used.

\subsection{\Ort}\label{sec:generatorBistection:ort}
The OSS Review Toolkit (\ort)~\cite{LinuxFoundation.ORT} is a community-based project for managing software components hosted by the Linux Foundation.
Unlike the others, \ort{} relies primarily on dynamic execution rather than static parsing and parsed files only to identify the appropriate project management tools.
For Python, it was running \texttt{pip}, \texttt{pipenv}, or \texttt{poetry} within virtual environments or executing \texttt{setup.py} (If the analyzeSetupPyInsecurely flag was set).
The actual metadata resolution is offloaded to the external \texttt{python-inspector}~\cite{AboutCode.2013} utility, which queries PyPi.
While executing code can lead to more precise component identification, it also poses security risks if the code is malicious.
Accordingly, a warning message is noted in the code and the documentation: \emph{``If "true", `python-inspector` resolves dependencies from setup.py files by executing them. This is a potential security risk.''}~\cite[line: 55 f.]{ort.pip.kt, ORT.Documentation.Poetry}.
For Java, \ort{} queried Gradle and Maven, to resolve dependencies and executed the gradlew.
Regarding Java SBT, \ort{} injects a custom SBT plugin and executes the application to identify the components.
For Go, the \texttt{go mod graph} and \texttt{go list} commands are used.
For PHP, it interacts with Composer.
For Rust, \texttt{cargo metadata} is used for detection.
For C, Conan (\texttt{conan info}, \texttt{conan graph info}) and Bazel (\texttt{bazel mod graph}, \texttt{bazel query}) are invoked.

\subsection{\SBOMTOOL{}}\label{sec:generatorBistection:sbomtool}
The Microsoft SBOM Tool~\cite{MS.SBOMTOOL} utilized the Microsoft Component Detection Library~\cite{Microsoft.Component.Detection} for identifying components.
For Python, \sbomtool{} supported Pip by executing \texttt{pip install --report}.
It also supported Poetry and Conda by statically parsing the management files. 
Additionally, as a legacy fallback method for pip environments, the tool executed the \texttt{setup.py} and parsed the requirements.txt.
For Java, Maven, and Gradle, parsing of manifest files is supported.
For Go, \sbomtool{} parsed the management files and used the \texttt{go list} command to identify components. 
PHP was not supported by \sbomtool{}.
For Rust, it supported cargo by parsing the management files. 
In addition to the documentation, \sbomtool{} supported C by parsing management files.
\section{Limitations}\label{dis:limitations}
When assessing our results, it is important to take the limitations of our approach into account.
These include threats to internal, external, and construct validity.

\subsection{Threats to Internal Validity}
We executed the SBOM generators with out-of-the-box settings to ensure ecological validity and comparability.
This approach is consistent with other studies in the Supply Composition Analysis (SCA) field~\cite{Shu.2025}.
However, further optimizations could increase completeness and correctness, so our results represent the minimum set of identified components.

\subsection{Threats to External Validity}
\textbf{Selection of SBOM Generators.}
We focused on \empirical{\num{5}} well-known, open-source tools. 
The tools are the most evaluated in academic literature and are recommended by tooling guidelines.
\Ort{} additionally documents its use by major industry partners, and \cdxgen{} is the tool developed by OWASP, the publisher of the CycloneDX standard, which we evaluated.
Consequently, it is likely that the investigated tools represent the state of the art and translate to the behavior of other SBOM generators in general.
However, there are other tools, particularly those specializing in individual programming languages or tools that are closed-source software, whose approaches we have not analyzed.

\textbf{Standard Conform Implementation.}
We implement our assessment tests in line with the standards of the programming languages.
This includes, for example, naming conventions for manifest files.
However, modifications might affect the detection rate.

\textbf{Transitive Depth.}
In addition, in compliance with the definition of the CRA~\cite{CRA.2024}, we decided to use only directly integrated components and refrain from testing transitive component detection.
Transitive component detection, however, is an important feature to reasonably argue about the supply chain of a software project.

\textbf{Dependency on the Environment Used.}
Build and runtime components are derived from the system being used.
How reliably they can be detected, therefore, depends on the system and may vary on other systems.
To provide a controlled, reproducible environment, we ran all tests in a Docker container.

\textbf{Assessment Scope.} 
Our assessment is based on the CIMs presented in Section~\ref{sec:dependencyInclusionTaxonomy}.
For each CIM, we provided at least one test per programming language, focusing on the standard implementations of that language.
However, other implementations of the same approach, such as additional package managers, may yield different detection outcomes.
Further, our scope does not include container environments.

\subsection{Threats to Construct Validity}

\textbf{Completeness of CIMs.}
Our CIMs represent a lower bound for the possible integration methods that actually exist.
While we have shown that the listed approaches exist and have security implications, we must not claim completeness, especially when accounting for language-specific idiosyncrasies.

\textbf{No focus on optional components.}
Some project management tools allow differentiating between required and optional components (e.g., components relevant only to developers).
In our tests, all components are required, and we do not evaluate how SBOM tools handle optional components.
\section{Discussion}\label{Discussion}
A fundamental prerequisite for mitigating software supply chain attacks using SBOMs is that an SBOM accurately represents the components of a software.
If it lists components that do not exist (false positives), it creates additional work, which reduces acceptance of the technology.
If components are missing (false negatives), a false sense of security is created.

Our results show that it is not trivial to have an accurate SBOM and that deviations can arise from two factors:
First, \emph{we do not have a shared understanding of what a component is} that should appear in an SBOM.
Secondly, \emph{SBOM generators are technically not capable of recognizing all types of components}.
We now discuss our results and cover (1) what constitutes a component and to what extent different conceptions of it vary (Section~\ref{dis:Component}), (2) why this lack of a common understanding has practical implications (Section~\ref{dis:PracticalImplications}), (3) to what extent SBOM generators can conceptually identify certain types of components (Section~\ref{dis:SDLC}), (4) why current blind spots arise (Section~\ref{dis:Accuracy}), and (5) what steps are necessary to ensure reliable SBOMs in the long term (Section~\ref{dis:WayForward}).

\subsection{What is a component?}\label{dis:Dependency}\label{dis:Component}
To evaluate the accuracy of SBOMs, a reliable ground truth of what should be included in an SBOM is necessary.
The first step in establishing such is to get a common understanding of what a component is.
We found that there is no such definition, neither in research nor between regulatory bodies.

Overall, there is no common understanding of what constitutes a component.
Prior research diverges on this question when assessing previous publications.
E.g., \cite{Balliu.2023}, \cite{Rabbi.2024}, or \cite{Cofano.2024} include every element considered by the project management tool, whereas \cite{Xiao.2025} and \cite{Wang.2026b} have a broader inclusion criteria.
Regulation does not provide a clear definition either, as even the most fundamental documents are not specific and, in some cases, contradict researchers' understanding.
Given this state of affairs, it is not reasonable to expect tool vendors and their customers to clearly articulate their expectations, and our observed results are, thus, hardly surprising.

Therefore, we sought to develop a definition acceptable to all parties.
To this end, we focused on one use case that government agencies worldwide recognize and commonly reference: identifying vulnerabilities in the software supply chains.
As a result, we defined a component as any part whose implementation is sourced from outside the corresponding project and shipped as executable code or executed at runtime, thereby potentially introducing vulnerabilities.

While we consider this definition well-rounded, we deliberately miss aspects of programs that could be considered components, such as external services, APIs, or containers.
However, this does not involve integrating any code into the application itself, and differentiating between components and services aligns with the Cyclonedx standard, which separates components and services (e.g., APIs and microservices) as well~\cite{ECMA.CycloneDX.2025}.
Thus, we invite criticism and constructive feedback but propose a definition that excludes them.
To solve this dilemma, SBOMs could add a separate section covering these elements, a question, and a problem space we leave open for future work.

\keyinsight{SBOMs require a common definition of a component to be listed in an SBOM, and we propose one in this paper based on software supply chain security (ref.~\Cref{sec:dependencyInclusionTaxonomy}).}

\subsection{Practical implications of the missing shared understanding}\label{dis:PracticalImplications}
The ambiguities surrounding what constitutes a component are not merely theoretical; they are also evident in our assessment, which has led to our reported false positives.
We found that in most cases where our components were detected, additional components were listed that we classified as false positives.
However, clear errors, such as incorrect version numbers, occurred only in a minority of those.
Rather, the tool vendors' understanding of what a component is differed from ours.
For example, \syft{} listed the gradle-wrapper as a component.

Those differences were not only between us and the tools, but also among the tools themselves.
For example, \trivy{} listed lock files as components with the type ``application'', while the others did not.
\Cdxgen{} listed the original notion for the version used in the manifest files as an additional component, while the others did not.
The \sbomtool{} did not consider Rust a component while the others did.
\Cdxgen{} contained the PYTHON\_EXCLUDED\_COMPONENTS~\cite[line: 384 ff.]{cdxgen.utils.js} specifying elements such as pip, setuptools, or conda that should not be listed as components.

SBOMs are intended to list components that are part of the supply chain~\cite{EO14028, CISA.2025, NIST.Def.SBOM.2026} to support cybersecurity efforts.
However, to build risk management on SBOMs, organizations need to know which components should appear in an SBOM.
But the divergence between the tools makes any such assessment impossible:
Is Rust not listed because it is not present, or because the tool determined that it is not a component?
Is pip not listed because it is not used, or because the generator filtered it out?
Without authoritative answers to these questions, organizations cannot determine the suitability of their toolchain, and thus failures in securing the software supply chain will be inevitable.

\keyinsight{Without a clear understanding of what a component is, no expectations against tools can be formulated, and thus, tools cannot be properly evaluated.}

\subsection{Components in the SDLC}\label{dis:SDLC}
Components can be introduced in different stages of the SDLC, as described in Section~\ref{taxonomy}: (1) As part of the source code, (2) as part of the build process, (3) dynamically during runtime.

While many regulatory bodies generally refer to the general term ``SBOM''~\cite{EO14028, CRA.2024}, we argue that the better approach is to follow the recommendation of the CISA and the BSI~\cite{CISA.2023b, TR-03183-2.2024} and clearly identify the type of SBOM generated (See Table~\ref{tbl:sbomtypes}).
As different type of SBOMs carry different implications of covered components:
If the source code serves as the basis, the SBOM describes a software project.
If the SBOM is generated as part of the build process, it describes the build artifacts and not the software project.
If the SBOM includes dynamically linked components, it describes a program in the context of a \emph{specific} system and cannot be transferred to another execution environment.

Our code analysis and test results have shown that the current focus is primarily on analyzing source code files with some capability for build artifacts.
Analysis of the runtime environment can be performed only in rare cases, such as when \cdxgen{} analyzes installed binaries and headers, and even then, only if the corresponding flag is set.
This inevitably will lead to blind spots in the software supply chain if no mitigation is deployed.

\keyinsight{The SBOM type is an essential indicator for the security perspective, as it implies which components an SBOM can reasonably be expected to include.}

\subsection{Component detection capabilities}\label{dis:Accuracy}
Using our reference implementations (\Cref{Benchmark}) to evaluate which component types the tools recognize and include in the generated SBOMs, we discovered that the tools share a core set of components they can detect, but also largely share the same blind spots.

\subsubsection{Detection of Managed Components}
The key finding is that tools can detect managed components with high reliability as they supported parsing package manager metadata files or running the package managers to obtain a list of components.
Since they are generated by project management tools and include the actual components loaded, they are guaranteed to be accurate.

This does, however, come with a clear drawback: these metadata files must be included in a project, and some tools rely exclusively on lock files.
Additionally, these metadata files need to be precise and complete, which may not always be the case, either.
Yu et al.~\cite{Yu.2024} reported that only 46\% of all \texttt{requirements.txt} files list explicit version numbers, implying a huge potential for blind spots.
 
Finally, what led to good results in our test may not always work in practice.
While GitHub recommends not excluding lockfiles~\cite{github.gitignore.templates}, there is no guarantee that they will actually be provided.
Tools that rely solely on lockfiles without a fallback method may therefore run into a dead end, possibly without the user noticing.
While we found that \cdxgen{} issues a quality warning if no lockfile is present, this was not the case for every generator.

\keyinsight{SBOM generators primarily rely on project management tools and their artifacts, with corresponding blind spots if these files are not provided.}

\subsubsection{Detection of Non-Managed Components}
In all tests outside of Managed Components, the tools were most successful for Java and Go.
Here, they benefited from the fact that Java provides standardized metadata files within the archives and that Go relies heavily on go.mod/go.sum files and cannot be compiled without them.
\Cdxgen{}, \syft{}, and \trivy{} successfully analyzed those files, thereby identifying our components.
However, especially for Go, the language's decision to rely exclusively on management files led to good results but arguably makes the corresponding tests not non-managed.

Going beyond this, only \cdxgen{} supported parsing program code, and even then, only with a specific flag, allowing it to find include statements.
This approach was particularly advanced in C, as \Cdxgen{} could search for include statements and detect installed libraries on the system, then link this information together.

Tools could also execute code, provided the correct flags were set.
For example, \cdxgen{} created a Python virtual environment and installed the specified components, while \ort{} offered the option to execute \texttt{setup.py} files.
This approach can further increase accuracy, but it does pose a security risk, which the tools recognize.
\ort{} has corresponding comments in the code and documentation~\cite[line: 55 f.]{ort.pip.kt, ORT.Documentation.Poetry} and \cdxgen{} a GitHub Issue raised by a user~\cite{Prabhu.2024}, which even led to a CVE~\cite{CVE.2024.50611}.
Given that SBOMs for larger projects may be part of CI/CD pipelines~\cite{Nocera.2024} and therefore run on strategically important servers, executing untrusted code is not always an option.

\keyinsight{While some tools go beyond parsing files and executing code, this approach poses a significant security risk.}

\subsubsection{Tool Philosophies}
In addition to strategies for identifying components, source code analysis has also revealed the different philosophies behind the tools.
However, the results of our tests do not always reflect the desired goal.
While our findings for \cdxgen{} reflect its philosophy that ``Some SBOM is better than no SBOM''~\cite[line: 14739 ff.]{cdxgen.utils.js}, they did not for others.
\Ort{}, for example, states that ``ORT's default philosophy is to analyze and scan everything it can find to build a complete picture of a repository and its dependencies''~\cite{ORT.Configuration.2025}.
However, our results show that \ort{} is the least complete while being the most accurate.
Additionally, if certain files that \ort{} expected for the analysis were not detected, the tool terminated the analysis with an error and did not produce an SBOM if the ``allowDynamicVersions'' option was not set.
This contrasted with the behavior of the other tools, which, when in doubt, generated an SBOM without components.
While not generating an SBOM clearly indicates issues, it prevents even a tentative supply chain analysis, whereas a generated but empty SBOM might mislead an inattentive user.

\keyinsight{The choice of tool affects the properties of the generated SBOM.}

\subsection{Are we there yet?}
For SBOMs to be an effective tool against supply chain attacks, they must accurately represent the components of a software product.

The goal of this work is to establish a common understanding of what a component is and demonstrate its implications if the definition is tested broadly across generators and programming languages.
However, our tests targeted a broad range of programming languages, and we did not test language specifics leading to only a high-level proposal on what components to list in an SBOM.
Determining which language-specific components are expected in an SBOM is a task for future work.
Finally, the SBOM generators need the technical capabilities to detect the expected types.
We observed clear blind spots with the evaluated SBOM generators.
While the tools examined are an important step in the right direction and make a significant contribution to supply chain security, their focus seems to be on source components, leaving significant gaps in Build and Runtime components.
Thus, we have to answer the titular question with \emph{not yet}.

\keyinsight{No, we do not have complete SBOMs yet.}

\subsection{Way Forward}\label{dis:WayForward}
Based on our discussion, three factors must be improved to make SBOM technology an effective response to supply chain attacks:
(1) A shared understanding of what a component is,
(2) A definition of what types of components can be expected in an SBOM, and
(3) the technical capabilities to identify those.

One of the main contributions of this paper is to provide a definition of a component and an initial categorization of the different types.
This is the first step toward strengthening SBOM technology.
However, we still need to establish it as a universally recognized definition.
Once we have a common understanding of what a component is and the types of components, we need to define which components we expect to find in an SBOM.
Although an initial answer might be ``all'', SBOMs are always generated in a certain context and often cannot identify all components.
Additionally, ``all'' is an unknown set and may differ across programming languages and setups.
The NTIA~\cite{NTIA.2021e} and the BSI~\cite{TR-03183-2.2024} have already taken an initial step by proposing SBOM types based on the SDLC.
Therefore, we recommend that an official matrix of SBOM use cases and SDLC generation points be created and filled with tangible examples of component inclusion for any programming language.
Otherwise, we are bound to miss components either in the SBOM itself or its corresponding list of known limitations.
As a first step towards this, we invite researchers and practitioners to submit suggestions to expand our test suite for security-related components, gradually broadening the assessment.
Finally, we also need tools with the technical capability to do so, and based on any testing of tools against the official examples, existing SBOM generators need to be improved or new ones developed.
\section{Conclusion}\label{Conclusion}
In this work, we first developed a broad definition of what we consider to be a component and derived multiple Component Inclusion Methods across the SDLC.
Subsequently, we developed an assessment that provides test cases for the \empirical{\num{6}} programming languages Python, Java, PHP, Go, Rust, and C, which integrate components in the identified ways.
This enabled us to generate SBOMs using the \empirical{\num{5}} SBOM generators \cdxgen{}, \syft{}, \trivy{}, \ort{}, and the \sbomtool{} and to evaluate whether they accurately recognized the integrated components.
To ensure the tools were running correctly, we also analyzed their capabilities using documentation and source code and compared them with the test results.

While components integrated by package managers were often recognized, components integrated during build or runtime revealed larger blind spots.
Most tools query the package managers or operate on metadata, such as manifest files or lockfiles.

Our overall conclusion is that although current tools are quite reliable at identifying managed components, large gaps remain.

We need to act on this observation and improve SBOMs further, starting with a fundamental discussion on their content to ensure a shared understanding.
Otherwise, we encourage organizations to rely on incomplete information to address supply chain attacks.
If we are unaware that vulnerable components are missing, we are not enhancing security; in the worst case, we are reducing it.

\section{Acknowledgements}
This work was funded by the Deutsche Forschungsgemeinschaft (DFG, German Research Foundation) under Germany's Excellence Strategy - EXC 2092 CASA - 390781972.

\clearpage
\bibliographystyle{IEEEtran}
\bibliography{bibliography}

@misc{link:pythonHistory,
url={https://docs.python.org/3/faq/general.html},
title={General Python FAQ},
urldate = {2026-01-10},
author={python.org},
year={2026}
}

@misc{link:golang,
url={https://go.dev/talks/2012/splash.article},
title={Go at Google: Language Design in the Service of Software Engineering},
lastaccessed = {2025-12-12},
author={Pike, Rob},
year={2012}
}

@misc{link:github:goNoBinaryPackage,
url={https://github.com/golang/go/issues/28152},
title={cmd/go: drop support for binary-only packages \#28152},
urldate = {2025-12-12},
author={Russ Cox},
year={2018}
}

@misc{link:googlesources:goNoBinaryPackage,
url={https://go-review.googlesource.com/c/go/+/165746},
title={cmd/go: drop support for binary-only packages},
urldate = {2025-12-12},
author={Jay Conrod},
year={2019}
}

@misc{javaUberJar,
url={https://support.sonatype.com/hc/en-us/articles/28958118202131-What-is-an-uber-jar},
title={What is an uber jar?},
urldate = {2025-12-12},
author={Sonatype Inc.},
year={2025}
}

@misc{javaPanama,
url={https://openjdk.org/projects/panama/},
title={Project Panama: Interconnecting JVM and native code},
urldate = {2026-01-10},
author={Oracle Corporation},
year={2026}
}

@misc{phpCFFI,
url={https://www.php.net/manual/en/book.ffi.php},
title={PHP Foreign Function Interface},
urldate = {2026-01-10},
author={The PHP Documentation Group},
year={2026}
}

@misc{link:codereusevuln,
url={https://thehackernews.com/2025/11/researchers-find-serious-ai-bugs.html},
title={Researchers Find Serious AI Bugs Exposing Meta, Nvidia, and Microsoft Inference Frameworks},
author={ Lakshmanan, Ravie},
urldate={2025-12-15},
year={2025}
}

@inproceedings{WiSonWWW2022,
author = {Wi, Seongil and Woo, Sijae and Whang, Joyce Jiyoung and Son, Sooel},
title = {HiddenCPG: Large-Scale Vulnerable Clone Detection Using Subgraph Isomorphism of Code Property Graphs},
year = {2022},
isbn = {9781450390965},
publisher = {Association for Computing Machinery},
address = {New York, NY, USA},
url = {https://doi.org/10.1145/3485447.3512235},
doi = {10.1145/3485447.3512235},
booktitle = {Proceedings of the ACM Web Conference 2022},
pages = {755–766},
numpages = {12},
location = {Virtual Event, Lyon, France},
series = {WWW '22}
}

@inproceedings{10.1145/3617555.3617872,
author = {Reid, David and Rahkema, Kristiina and Walden, James},
title = {Large Scale Study of Orphan Vulnerabilities in the Software Supply Chain},
year = {2023},
isbn = {9798400703751},
publisher = {Association for Computing Machinery},
address = {New York, NY, USA},
url = {https://doi.org/10.1145/3617555.3617872},
doi = {10.1145/3617555.3617872},
booktitle = {Proceedings of the 19th International Conference on Predictive Models and Data Analytics in Software Engineering},
pages = {22–32},
numpages = {11},
location = {San Francisco, CA, USA},
series = {PROMISE 2023}
}

@misc{web_technologies_overview,
url={https://w3techs.com/technologies/overview/programming_language},
title={Usage statistics of server-side programming languages for websites},
author={W3Techs},
urldate={2026-01-10},
year={2026}
}

@misc{java_usage,
url={https://www.secondtalent.com/resources/domain-java-statistics/},
title={Java Statistics: Adoption, Usage, and Future Trends},
author={Second Talent},
urldate={2026-01-10},
year={2025}
}

@misc{python_usage,
url={https://www.tiobe.com/tiobe-index/},
title={TIOBE Index for January 2026},
author={TIOBE Software BV},
urldate={2026-01-10},
year={2026}
}

@misc{go_documentation,
url={https://go.dev/doc/},
title={Go Documentation},
author={Google},
urldate={2026-01-10},
year={2026}
}

@misc{java_jvm,
url={https://docs.oracle.com/en/java/javase/25/vm/java-virtual-machine-technology-overview.html},
title={Java Virtual Machine Technology Overview},
author={Oracle},
urldate={2026-01-10},
year={2026}
}

@misc{java_first_release,
url={https://web.archive.org/web/20070310235103/http://www.sun.com/smi/Press/sunflash/1996-01/sunflash.960123.10561.xml},
title={JAVASOFT SHIPS JAVA 1.0},
author={Sun microsystems},
urldate={2026-01-10},
year={1996}
}

@misc{php.history,
 author = {The PHP Documentation Group},
 date = {2026},
 title = {History of PHP},
 url = {https://www.php.net/manual/en/history.php.php},
 lastaccessed = {2026-01-10},
 year = {2026}
}

@misc{cdxgen.index.js,
 author = {cdxgen},
 date = {2026},
 title = {index.js},
 url = {https://github.com/cdxgen/cdxgen/blob/master/lib/cli/index.js},
 lastaccessed = {2026-01-13},
 year = {2026}
}

@misc{cdxgen.utils.js,
 author = {cdxgen},
 date = {2026},
 title = {utils.js},
 url = {https://github.com/cdxgen/cdxgen/blob/master/lib/helpers/utils.js},
 lastaccessed = {2026-01-13},
 year = {2026}
}

@misc{ort.pip.kt,
 author = {The Linux Foundation},
 date = {2026},
 title = {Pip.kt},
 url = {https://github.com/oss-review-toolkit/ort/blob/main/plugins/package-managers/python/src/main/kotlin/Pip.kt},
 lastaccessed = {2026-01-13},
 year = {2026}
}

@misc{AboutCode.2013,
 author = {AboutCode},
 date = {2013},
 title = {python-inspector - inspect Python package dependencies and metadata},
 url = {https://github.com/aboutcode-org/python-inspector},
 lastaccessed = {2025-10-30},
 institution = {AboutCode},
 year = {2013}
}

@misc{AdvancedSecurity.GH-SBOM,
 author = {{Advanced Security}},
 date = {2023},
 title = {gh-sbom},
 url = {https://github.com/advanced-security/gh-sbom},
 lastaccessed = {2025-07-21},
 institution = {{Advanced Security}},
 year = {2023}
}

@article{Ahmed.2024,
 author = {Ahmed, Akinsola and Abdullah, Akinde},
 year = {2024},
 title = {Enhancing Software Supply Chain Resilience: Strategy for Mitigating Software Supply Chain Security Risks and Ensuring Security Continuity in Development Lifecycle},
 pages = {01--18},
 volume = {15},
 number = {1/2},
 issn = {22297103},
 journal = {International Journal on Soft Computing},
 doi = {10.5121/ijsc.2024.15201}
}

@misc{Alrich.2020,
 author = {Alrich, Tom},
 date = {2020},
 title = {The naming problem},
 url = {https://tomalrichblog.blogspot.com/2020/11/the-naming-problem.html},
 lastaccessed = {2026-03-09},
 year = {2020}
}

@misc{Anchore.Docs.2026,
 author = {Anchore},
 date = {2026},
 title = {Capabilites},
 url = {https://oss.anchore.com/docs/capabilities/},
 lastaccessed = {2026-04-14},
 year = {2026}
}

@misc{Atlassian.2025,
 author = {{Atlassian Pty Ltd.}},
 date = {2025},
 title = {Git ignore},
 url = {https://www.atlassian.com/git/tutorials/saving-changes/gitignore},
 lastaccessed = {2025-11-12},
 year = {2025}
}

@misc{Atom.2026,
 author = {AppThreat},
 date = {2026},
 title = {Atom},
 url = {https://github.com/AppThreat/atom},
 lastaccessed = {2026-04-08},
 year = {2026}
}

@misc{Autodesk.2020,
 author = {Autodesk},
 date = {2020},
 title = {MAXScript exploit {\textquotedbl}PhysXPluginMfx{\textquotedbl} in Autodesk 3ds Max software},
 url = {https://www.autodesk.com/trust/security-advisories/adsk-sa-2020-0005},
 lastaccessed = {2026-03-11},
 year = {2020}
}

@article{Balliu.2023,
 author = {Balliu, Musard and Baudry, Benoit and Bobadilla, Sofia and Ekstedt, Mathias and Monperrus, Martin and Ron, Javier and Sharma, Aman and Skoglund, Gabriel and Soto-Valero, C{\'e}sar and Wittlinger, Martin},
 year = {2023},
 title = {Challenges of Producing Software Bill of Materials for Java},
 pages = {12--23},
 volume = {21},
 number = {6},
 issn = {1540-7993},
 journal = {IEEE Security {\&} Privacy},
 doi = {10.1109/MSEC.2023.3302956}
}

@incollection{Benedetti.2025b,
 author = {Benedetti, Giacomo and Cofano, Serena and Brighente, Alessandro and Conti, Mauro},
 title = {The Impact of SBOM Generators on Vulnerability Assessment in Python: A Comparison and a Novel Approach},
 pages = {487--509},
 volume = {15826},
 publisher = {{Springer Nature Switzerland}},
 isbn = {978-3-031-95763-5},
 series = {Lecture Notes in Computer Science},
 editor = {Fischlin, Marc and Moonsamy, Veelasha},
 booktitle = {Applied Cryptography and Network Security},
 year = {2025},
 address = {Cham},
 doi = {10.1007/978-3-031-95764-2{\textunderscore }19}
}

@article{Bi.2024,
 author = {Bi, Tingting and Xia, Boming and Xing, Zhenchang and Lu, Qinghua and Zhu, Liming},
 year = {2024},
 title = {On the Way to SBOMs: Investigating Design Issues and Solutions in Practice},
 pages = {1--25},
 volume = {33},
 number = {6},
 issn = {1049-331X},
 journal = {ACM Transactions on Software Engineering and Methodology},
 doi = {10.1145/3654442}
}

@inproceedings{Bifolco.2024,
 author = {Bifolco, Daniele and Nocera, Sabato and Romano, Simone and {Di Penta}, Massimiliano and Francese, Rita and Scanniello, Giuseppe},
 title = {On the Accuracy of GitHub's Dependency Graph},
 pages = {242--251},
 publisher = {ACM},
 isbn = {9798400717017},
 booktitle = {Proceedings of the 28th International Conference on Evaluation and Assessment in Software Engineering (EASE)},
 year = {2024},
 doi = {10.1145/3661167.3661175},
 address = {Salerno Italy}
}

@misc{Brabandere.2025,
 author = {de Brabandere, Jo},
 date = {2025},
 title = {Demystifying the Software Bill of Materials (SBOM) -- the key to software supply chain security},
 url = {https://cybersecuritycoalition.be/resource/demystifying-the-software-bill-of-materials-sbom-the-key-to-software-supply-chain-security/},
 lastaccessed = {2026-03-09},
 year = {2025}
}

@misc{CDX.Conan,
 author = {CycloneDX},
 date = {2021},
 title = {CycloneDX Conan},
 url = {https://github.com/CycloneDX/cyclonedx-conan},
 lastaccessed = {2025-07-21},
 institution = {CycloneDX},
 year = {2021}
}

@misc{CDX.DotNet,
 author = {CycloneDX},
 date = {2018},
 title = {CycloneDX module for .NET},
 url = {https://github.com/CycloneDX/cyclonedx-dotnet},
 lastaccessed = {2025-07-21},
 institution = {CycloneDX},
 year = {2018}
}

@misc{CDX.MavenPlugin,
 author = {CycloneDX},
 date = {2017},
 title = {CycloneDX Maven Plugin},
 url = {https://github.com/CycloneDX/cyclonedx-maven-plugin},
 institution = {CycloneDX},
 year = {2017}
}

@misc{CDX.npm,
 author = {CycloneDX},
 date = {2022},
 title = {CycloneDX SBOM for npm},
 url = {https://www.npmjs.com/package/@cyclonedx/cyclonedx-npm},
 lastaccessed = {2025-07-21},
 institution = {CycloneDX},
 year = {2022}
}

@misc{CDX.PHP,
 author = {CycloneDX},
 date = {2019},
 title = {CycloneDX PHP Composer Plugin},
 url = {https://github.com/CycloneDX/cyclonedx-php-composer},
 lastaccessed = {2025-07-21},
 institution = {CycloneDX},
 year = {2019}
}

@misc{CDX.Python,
 author = {CycloneDX},
 date = {2018},
 title = {CycloneDX Python SBOM Generation Tool},
 url = {https://github.com/CycloneDX/cyclonedx-python},
 lastaccessed = {2025-07-21},
 institution = {CycloneDX},
 year = {2018}
}

@misc{cdxgen,
 author = {CycloneDX},
 date = {2017},
 title = {CycloneDX Generator (cdxgen)},
 url = {https://github.com/CycloneDX/cdxgen},
 lastaccessed = {2025-07-21},
 institution = {CycloneDX},
 year = {2017}
}

@misc{Chainguard,
 author = {Chainguard},
 date = {2025},
 title = {Chainguard Platform},
 url = {https://edu.chainguard.dev/open-source/sbom/},
 lastaccessed = {2025-07-21},
 institution = {Chainguard},
 year = {2025}
}

@misc{CISA.2022,
 author = {{Cybersecurity and Infrastructure Security Agency}},
 date = {2022},
 title = {Review of the December 2021 Log4j Event},
 url = {https://www.cisa.gov/sites/default/files/publications/CSRB-Report-on-Log4-July-11-2022_508.pdf},
 lastaccessed = {2025-07-22},
 institution = {{Cybersecurity and Infrastructure Security Agency}},
 year = {2022}
}

@misc{CISA.2023b,
 author = {{Cybersecurity and Infrastructure Security Agency}},
 date = {2023},
 title = {Types of Software Bill of Material (SBOM) Documents},
 url = {https://www.cisa.gov/resources-tools/resources/types-software-bill-materials-sbom},
 lastaccessed = {2025-03-29},
 institution = {{Cybersecurity and Infrastructure Security Agency}},
 year = {2023}
}

@misc{CISA.2025,
 author = {{Cybersecurity and Infrastructure Security Agency}},
 date = {2025},
 title = {A Shared Vision of Software Bill of Materials (SBOM) for Cybersecurity},
 url = {https://www.cisa.gov/resources-tools/resources/shared-vision-software-bill-materials-sbom-cybersecurity},
 lastaccessed = {2025-09-04},
 institution = {{Cybersecurity and Infrastructure Security Agency}},
 year = {2025}
}

@misc{CISA.ShaiHulud.2025,
 author = {{Cybersecurity and Infrastructure Security Agency}},
 date = {2025},
 title = {Widespread Supply Chain Compromise Impacting npm Ecosystem: Alert},
 url = {https://www.cisa.gov/news-events/alerts/2025/09/23/widespread-supply-chain-compromise-impacting-npm-ecosystem},
 lastaccessed = {2025-12-19},
 institution = {{Cybersecurity and Infrastructure Security Agency}},
 year = {2025}
}

@misc{CISA.Stuxnet.2013,
 author = {{Cybersecurity and Infrastructure Security Agency}},
 date = {2013},
 title = {Siemens SIMATIC STEP 7 DLL Vulnerability},
 url = {https://www.cisa.gov/news-events/ics-advisories/icsa-12-205-02},
 lastaccessed = {2026-03-11},
 year = {2013}
}

@inproceedings{Cofano.2024,
 author = {Cofano, Serena and Benedetti, Giacomo and Dell'Amico, Matteo},
 title = {SBOM Generation Tools in the Python Ecosystem: an In-Detail Analysis},
 pages = {427--434},
 publisher = {IEEE},
 isbn = {979-8-3315-0620-9},
 booktitle = {2024 IEEE 23rd International Conference on Trust, Security and Privacy in Computing and Communications (TrustCom)},
 year = {2024},
 doi = {10.1109/TrustCom63139.2024.00077},
 address = {Sanya, China}
}

@misc{Composer.2025b,
 author = {Adermann, Nils and Boggiano, Jordi},
 date = {2025},
 title = {Composer: Versions and constraints},
 url = {https://getcomposer.org/doc/articles/versions.md},
 lastaccessed = {2025-10-02},
 year = {2025}
}

@misc{Composer.2025c,
 author = {Adermann, Nils and Boggiano, Jordi},
 date = {2025},
 title = {Composer: Basic usage},
 url = {https://getcomposer.org/doc/01-basic-usage.md},
 lastaccessed = {2025-11-09},
 year = {2025}
}

@misc{Composer.2026,
 author = {Adermann, Nils and Boggiano, Jordi},
 date = {2026},
 title = {Composer: A Dependency Manager for PHP},
 url = {https://getcomposer.org/},
 lastaccessed = {2026-01-11},
 year = {2026}
}

@misc{Conan.2026,
 author = {JFrog},
 date = {2026},
 title = {Conan 2},
 url = {https://conan.io/},
 lastaccessed = {2026-04-11},
 year = {2026}
}

@misc{Conan2.versionranges,
 author = {JFrog},
 date = {2026},
 title = {Conan2 Documentation: Version ranges},
 url = {https://docs.conan.io/2/tutorial/versioning/version_ranges.html},
 lastaccessed = {2026-04-11},
 year = {2026}
}

@misc{ContrastSecurity.jbom,
 author = {{Contrast Security}},
 date = {2021},
 title = {jbom},
 url = {https://github.com/eclipse-jbom/jbom},
 lastaccessed = {2025-07-21},
 institution = {{Contrast Security}},
 year = {2021}
}

@misc{corbet.2025,
 author = {corbet},
 date = {2025},
 title = {The (successful) end of the kernel Rust experiment},
 url = {https://lwn.net/Articles/1049831/},
 lastaccessed = {2026-04-02},
 year = {2025}
}

@misc{CRA.2024,
 author = {{European Parliament and the council of the european union}},
 title = {Cyber Resilience Act: CRA},
 year = {2024},
 url = {https://eur-lex.europa.eu/legal-content/EN/TXT/?uri=CELEX:32024R2847},
 urldate = {2025-03-17}
}

@inproceedings{Crawford.2023,
 author = {Crawford, Alex and Yakubovich, Eugene and Szumski, Rob},
 title = {Enforcing SBOMs through the Linux kernel with eBPF and IMA},
 pages = {77--78},
 publisher = {ACM},
 isbn = {9798400702631},
 editor = {Torres-Arias, Santiago and Melara, Marcela and Simon, Laurent and Vasilakis, Nikos and Moriarty, Kathleen},
 booktitle = {Proceedings of the 2023 Workshop on Software Supply Chain Offensive Research and Ecosystem Defenses},
 year = {2023},
 doi = {10.1145/3605770.3625206},
 address = {Copenhagen Denmark}
}

@misc{CVE.2024.50611,
 author = {{Information Technology Laboratory}},
 date = {2024},
 title = {CVE-2024-50611},
 url = {https://nvd.nist.gov/vuln/detail/CVE-2024-50611},
 lastaccessed = {2026-04-14},
 year = {2024}
}

@misc{CVE-2021-25290,
 author = {{National Institute of Standards and Technology (NIST)}},
 date = {2021},
 title = {CVE-2021-25290 Detail},
 url = {https://nvd.nist.gov/vuln/detail/cve-2021-25290},
 lastaccessed = {2026-03-11},
 year = {2021}
}

@misc{CycloneDX,
 author = {{Open Web Application Security Project}},
 date = {2025},
 title = {CycloneDX Specification Overview},
 url = {https://cyclonedx.org/specification/overview/},
 lastaccessed = {2025-08-20},
 institution = {{Open Web Application Security Project}},
 year = {2025}
}

@misc{CycloneDX.CLI.2020,
 author = {CycloneDX},
 date = {2020},
 title = {CycloneDX CLI},
 url = {https://github.com/CycloneDX/cyclonedx-cli},
 lastaccessed = {2025-10-09},
 year = {2020}
}

@misc{CycloneDX.PythonLibrary.2026,
 author = {{Open Web Application Security Project}},
 date = {2026},
 title = {cyclonedx-python-lib},
 url = {https://github.com/CycloneDX/cyclonedx-python-lib},
 lastaccessed = {2026-03-09},
 year = {2026}
}

@misc{CycloneDX.PythonLibrary.Docs.2026,
 author = {{Open Web Application Security Project}},
 date = {2026},
 title = {CycloneDX Python Library: Docs},
 url = {https://cyclonedx-python-library.readthedocs.io/},
 lastaccessed = {2026-09-03},
 year = {2026}
}

@inproceedings{Dalia.2024,
 author = {Dalia, Gregorio and Visaggio, Corrado Aaron and {Di Sorbo}, Andrea and Canfora, Gerardo},
 title = {SBOM Ouverture: What We Need and What We Have},
 pages = {1--9},
 publisher = {ACM},
 isbn = {9798400717185},
 booktitle = {Proceedings of the 19th International Conference on Availability, Reliability and Security},
 year = {2024},
 doi = {10.1145/3664476.3669975},
 address = {Vienna Austria}
}

@misc{Ebay.2022,
 author = {Ebay},
 date = {2022},
 title = {SBOM-Scorecard},
 url = {https://github.com/eBay/sbom-scorecard},
 lastaccessed = {2025-03-29},
 year = {2022}
}

@misc{ECMA.CycloneDX.2025,
 year = {2025},
 title = {CycloneDX Bill of materials specification},
 url = {https://ecma-international.org/wp-content/uploads/ECMA-424_2nd_edition_december_2025.pdf},
 urldate = {2026-03-09},
 number = {ECMA-424},
 author = {{European Computer Manufacturers Association}}
}

@misc{EO14028,
 author = {{President of the United States}},
 title = {Executive Order 14028: Improving the Nation's Cybersecurity},
 year = {2021},
 url = {https://www.federalregister.gov/documents/2021/05/17/2021-10460/improving-the-nations-cybersecurity},
 urldate = {2025-03-19},
 pages = {26633--26647},
 volume = {86},
 number = {93},
 publisher = {{National Archives and Records Administration}},
 editor = {{National Archives and Records Administration}},
 booktitle = {Federal Register}
}

@misc{Festo.2023,
 author = {{Festo SE {\&} Co. KG}},
 date = {2023},
 title = {CycloneDX Editor/Validator},
 url = {https://github.com/Festo-se/cyclonedx-editor-validator},
 lastaccessed = {2025-10-09},
 year = {2023}
}

@inproceedings{Fischer.2017,
 author = {Fischer, Felix and Bottinger, Konstantin and Xiao, Huang and Stransky, Christian and Acar, Yasemin and Backes, Michael and Fahl, Sascha},
 title = {Stack Overflow Considered Harmful? The Impact of Copy{\&}Paste on Android Application Security},
 pages = {121--136},
 publisher = {IEEE},
 isbn = {978-1-5090-5533-3},
 booktitle = {2017 IEEE Symposium on Security and Privacy (SP)},
 year = {2017},
 doi = {10.1109/SP.2017.31},
 address = {San Jose, CA, USA}
}

@misc{FOSSA,
 author = {{FOSSA Inc.}},
 date = {2025},
 title = {FOSSA},
 url = {https://fossa.com/},
 lastaccessed = {2025-07-21},
 institution = {{FOSSA Inc.}},
 year = {2025}
}

@inproceedings{Garcia.2025,
 author = {Garcia, Derek and Mirakorhli, Mehdi Tarrit and Dillon, Schuyler and Laporte, Kevin and Morrison, Matthew and Lu, Henry and Koscinski, Viktoria and Enoch, Christopher and Fazelnia, Mohamad and Chen, Roger},
 title = {A Landscape Study of Open-Source Tools for Software Bill of Materials (SBOM) and Supply Chain Security},
 pages = {37--45},
 publisher = {IEEE},
 isbn = {979-8-3315-1468-6},
 booktitle = {2025 IEEE/ACM 3rd International Workshop on Software Vulnerability Management (SVM)},
 year = {2025},
 doi = {10.1109/SVM66695.2025.00010},
 address = {Ottawa, ON, Canada}
}

@misc{GitHub.2010,
 author = {{GitHub, Inc.}},
 date = {2010},
 title = {A collection of .gitignore templates},
 url = {https://github.com/github/gitignore},
 lastaccessed = {2025-11-12},
 year = {2010}
}

@misc{github.gitignore.templates,
 author = {{GitHub, Inc.}},
 date = {2010},
 title = {gitignore},
 url = {https://github.com/github/gitignore},
 lastaccessed = {2026-04-14},
 year = {2010}
}

@misc{GitHubDependencyGraph,
 author = {{GitHub, Inc.}},
 date = {2025},
 title = {About the dependency graph},
 url = {https://docs.github.com/en/code-security/supply-chain-security/understanding-your-software-supply-chain/about-the-dependency-graph},
 lastaccessed = {2025-04-19},
 institution = {{GitHub, Inc.}},
 year = {2025}
}

@misc{GNU.OptionsForLinking.2026,
 author = {{Free Software Foundation, Inc.}},
 date = {2026},
 title = {GCC Online Docs: 3.16 Options for Linking},
 url = {https://gcc.gnu.org/onlinedocs/gcc/Link-Options.html},
 lastaccessed = {2026-03-10},
 year = {2026}
}

@misc{Go.2025b,
 author = {Google},
 date = {2025},
 title = {The Go Programming Language Specification},
 url = {https://tip.golang.org/ref/spec},
 lastaccessed = {2025-10-02},
 year = {2025}
}

@misc{Go.2025c,
 author = {Google},
 date = {2025},
 title = {Go: Plugin},
 url = {https://pkg.go.dev/plugin},
 lastaccessed = {2025-10-01},
 year = {2025}
}

@misc{go.faq,
 author = {Google},
 date = {2026},
 title = {GO FAQs},
 url = {https://go.dev/doc/faq},
 lastaccessed = {2026-01-10},
 year = {2026}
}

@misc{Go.mod.reference,
 author = {Google},
 date = {2026},
 title = {go.mod file reference},
 url = {https://go.dev/doc/modules/gomod-ref},
 lastaccessed = {2026-01-11},
 year = {2026}
}

@misc{Go.mvs,
 author = {Google},
 date = {2026},
 title = {Go Modules Reference: Minimal version selection (MVS)},
 url = {https://go.dev/ref/mod\#minimal-version-selection},
 lastaccessed = {2026-01-11},
 year = {2026}
}

@misc{Go.Wiki.2026,
 author = {Google},
 date = {2026},
 title = {Go Wiki: Go Modules},
 url = {https://go.dev/wiki/Modules},
 lastaccessed = {2026-03-09},
 year = {2026}
}

@misc{Gradle.2025b,
 author = {{Gradle, Inc.}},
 date = {2025},
 title = {Gradle: Declaring Versions and Ranges},
 url = {https://docs.gradle.org/current/userguide/dependency_versions.html},
 lastaccessed = {2025-10-02},
 year = {2025}
}

@misc{Gradle.2026,
 author = {{Gradle, Inc.}},
 date = {2026},
 title = {Gradle},
 url = {https://gradle.org/},
 lastaccessed = {2026-01-11},
 year = {2026}
}

@misc{Gradle.build.process,
 author = {{Gradle, Inc.}},
 date = {2026},
 title = {Gradle Build Lifecycle},
 url = {https://docs.gradle.org/current/userguide/build_lifecycle_intermediate.html},
 lastaccessed = {2026-01-11},
 year = {2026}
}

@inproceedings{Halbritter.2024,
 author = {Halbritter, Andreas and Merli, Dominik},
 title = {Accuracy Evaluation of SBOM Tools for Web Applications and System-Level Software},
 pages = {1--9},
 publisher = {ACM},
 isbn = {9798400717185},
 booktitle = {Proceedings of the 19th International Conference on Availability, Reliability and Security},
 year = {2024},
 doi = {10.1145/3664476.3670926},
 address = {Vienna Austria}
}

@misc{Harding.2025,
 author = {Harding, William},
 date = {2025},
 title = {AI Copilot Code Quality: Evaluating 2024's Increased Defect Rate via Code Quality Metrics},
 url = {https://www.gitclear.com/ai_assistant_code_quality_2025_research},
 lastaccessed = {2026-03-10},
 institution = {GitClear},
 year = {2025}
}

@misc{Harrison.2023,
 author = {Harrison, Anthony},
 date = {2023},
 title = {lib4sbom},
 url = {https://github.com/anthonyharrison/lib4sbom},
 lastaccessed = {2025-10-09},
 year = {2023}
}

@misc{Harrison.2023b,
 author = {Harrison, Anthony},
 date = {2023},
 title = {SBOMAUDIT},
 url = {https://github.com/anthonyharrison/sbomaudit},
 lastaccessed = {2025-10-09},
 year = {2023}
}

@misc{Heartbleed.NVD.2014,
 author = {{Information Technology Laboratory}},
 year = {2014},
 title = {CVE-2014-0160},
 url = {https://nvd.nist.gov/vuln/detail/cve-2014-0160},
 lastaccessed = {2025-04-27}
}

@misc{Interlynk.2023,
 author = {Interlynk},
 date = {2023},
 title = {sbomqs: The Comprehensive SBOM Quality {\&} Compliance Tool},
 url = {https://github.com/interlynk-io/sbomqs},
 lastaccessed = {2025-09-03},
 institution = {Interlynk},
 year = {2023}
}

@inproceedings{Islam.2017,
 author = {Islam, Md Rakibul and Zibran, Minhaz F. and Nagpal, Aayush},
 title = {Security Vulnerabilities in Categories of Clones and Non-Cloned Code: An Empirical Study},
 pages = {20--29},
 publisher = {IEEE},
 isbn = {978-1-5090-4039-1},
 booktitle = {2017 ACM/IEEE International Symposium on Empirical Software Engineering and Measurement (ESEM)},
 year = {2017},
 doi = {10.1109/ESEM.2017.9},
 address = {Toronto, ON}
}

@misc{ISO.19770.2015,
 year = {2015},
 title = {Information technology. Software asset management. Software identification tag},
 url = {https://www.iso.org/standard/65666.html},
 urldate = {2025-08-20},
 volume = {35.080 Software. Programme},
 number = {ISO/IEC 19770-2:2015},
 author = {{International Organization for Standardization (ISO)}}
}

@misc{JetBrains.2023,
 author = {{JetBrains s.r.o.}},
 date = {2023},
 title = {State of Developer Ecosystem Report 2023},
 url = {https://www.jetbrains.com/lp/devecosystem-2023/},
 lastaccessed = {2025-12-11},
 year = {2023}
}

@misc{JFrog,
 author = {JFrog},
 date = {2021},
 title = {Build-Info-Go},
 url = {https://github.com/jfrog/build-info-go},
 lastaccessed = {2025-07-21},
 institution = {JFrog},
 year = {2021}
}

@article{Kagzmandere.2024,
 author = {Ka{\u{g}}{\i}zmandere, {\"O}mercan and Arslan, Halil},
 year = {2024},
 title = {Vulnerability analysis based on SBOMs: A model proposal for automated vulnerability scanning for CI/CD pipelines},
 pages = {33--42},
 volume = {13},
 number = {2},
 journal = {International Journal of Information Security Science},
 doi = {10.55859/ijiss.1455039}
}

@inproceedings{Kawaguchi.2024,
 author = {Kawaguchi, Nobutaka and Hart, Charlie},
 title = {On the Deployment Control and Runtime Monitoring of Containers Based on Consumer Side SBOMs},
 pages = {1022--1025},
 publisher = {IEEE},
 isbn = {979-8-3503-0457-2},
 booktitle = {2024 IEEE 21st Consumer Communications {\&} Networking Conference (CCNC)},
 year = {2024},
 doi = {10.1109/CCNC51664.2024.10454654},
 address = {Las Vegas, NV, USA}
}

@article{Kim.2018,
 author = {Kim, Seulbae and Lee, Heejo},
 year = {2018},
 title = {Software systems at risk: An empirical study of cloned vulnerabilities in practice},
 pages = {720--736},
 volume = {77},
 issn = {01674048},
 journal = {Computers {\&} Security},
 doi = {10.1016/j.cose.2018.02.007}
}

@article{Kong.2025,
 author = {Kong, Xianglong and Zhuo, Hangyi and Miao, Xinyuan and Huang, Wei and Du, Jiayu},
 year = {2025},
 title = {SBOM generation based on code-level external component trees},
 pages = {45277},
 volume = {15},
 number = {1},
 journal = {Scientific reports},
 doi = {10.1038/s41598-025-29762-0}
}

@misc{Kubernetes.BOM,
 author = {Kubernetes},
 date = {2021},
 title = {Kubernetes Sigs Bom},
 url = {https://github.com/kubernetes-sigs/bom},
 lastaccessed = {2025-07-21},
 institution = {Kubernetes},
 year = {2021}
}

@inproceedings{Li.2024b,
 author = {Li, Hongyu and Guo, Liwei and Yang, Yexuan and Wang, Shangguang and Xu, Mengwei},
 title = {An Empirical Study of Rust-for-Linux: The Success, Dissatisfaction, and Compromise},
 pages = {425--443},
 booktitle = {2024 USENIX Annual Technical Conference (USENIX ATC 24)},
 year = {2024}
}

@misc{LinuxFoundation.ORT,
 author = {{The Linux Foundation}},
 date = {2017},
 title = {OSS Review Toolkit (ORT)},
 url = {https://github.com/oss-review-toolkit/ort},
 lastaccessed = {2025-07-21},
 institution = {{The Linux Foundation}},
 year = {2017}
}

@inproceedings{ManziMuneza.2025,
 author = {{Manzi Muneza}, A. Redempta and Keefe, Aidan and O'Donoghue, Eric and Izurieta, Clemente and Reinhold, Ann Marie},
 title = {SBOM Generation Tools and Formats Affect Compliance with US Standard},
 pages = {81--88},
 publisher = {IEEE},
 isbn = {979-8-3315-3749-4},
 booktitle = {2025 IEEE International Conference on Software Analysis, Evolution and Reengineering - Companion (SANER-C)},
 year = {2025},
 doi = {10.1109/SANER-C66551.2025.00019},
 address = {Montreal, QC, Canada}
}

@misc{Mead.2024,
 author = {Mead, Nancy R. and Woody, Carol and Hissam, Scott},
 date = {2024},
 title = {Measurement Challenges in Software Assurance and Supply Chain Risk Management},
 url = {https://www.sei.cmu.edu/blog/measurement-challenges-in-software-assurance-and-supply-chain-risk-management/},
 lastaccessed = {2025-08-20},
 institution = {{Carnegie Mellon University}},
 year = {2024}
}

@misc{Mens.2024,
 author = {Mens, Tom and Decan, Alexandre},
 date = {2024},
 title = {An Overview and Catalogue of Dependency Challenges in Open Source Software Package Registries},
 publisher = {arXiv},
 doi = {10.48550/arXiv.2409.18884},
 year = {2024}
}

@inproceedings{Merigala.2024,
 author = {Merigala, Joseph and Kumar, Vivek and Gujjarlapudi, Jashuva and Gupta, Manas and Kumar, Athmakuri Satish},
 title = {Analysis of Supply Chain Attacks in Open-Source Software and Mitigation Strategies},
 pages = {1--5},
 publisher = {IEEE},
 isbn = {979-8-3503-7988-4},
 booktitle = {2024 5th International Conference on Communication, Computing {\&} Industry 6.0 (C2I6)},
 year = {2024},
 doi = {10.1109/C2I663243.2024.10895136},
 address = {Bengaluru, India}
}

@misc{Meyers.2023,
 author = {Meyers, John Speed},
 date = {2023},
 title = {A purl of wisdom on SBOMs and vulnerabilities},
 url = {https://www.chainguard.dev/unchained/a-purl-of-wisdom-on-sboms-and-vulnerabilities},
 lastaccessed = {2026-03-09},
 institution = {Chainguard},
 year = {2023}
}

@misc{Microsoft.Component.Detection,
 author = {Microsoft},
 date = {2026},
 title = {Component Detection},
 url = {https://github.com/microsoft/component-detection},
 lastaccessed = {2026-04-08},
 year = {2026}
}

@misc{MITRE.Glossary,
 author = {{MITRE Corporation}},
 date = {2026},
 title = {CVE: Glossary},
 url = {https://www.cve.org/ResourcesSupport/Glossary},
 lastaccessed = {2026-03-15},
 institution = {{MITRE Corporation}},
 year = {2026}
}

@misc{Moderne.Rewrite,
 author = {Moderne},
 date = {2025},
 title = {OpenRewrite},
 url = {https://docs.openrewrite.org/recipes/java/dependencies/softwarebillofmaterials},
 lastaccessed = {2025-07-21},
 institution = {Moderne},
 year = {2025}
}

@misc{Mozilla.2017,
 author = {{Mozilla Corporation}},
 date = {2017},
 title = {Put your Trust in Rust -- Shipping Now in Firefox},
 url = {https://blog.mozilla.org/en/firefox/put-trust-rust-shipping-now-firefox/},
 lastaccessed = {2026-04-02},
 year = {2017}
}

@misc{MS.SBOMTOOL,
 author = {Microsoft},
 date = {2022},
 title = {Microsoft SBOM Tool},
 url = {https://github.com/microsoft/sbom-tool},
 lastaccessed = {2025-07-21},
 institution = {Microsoft},
 year = {2022}
}

@misc{Naveen.2025,
 author = {Naveen, Nathan},
 date = {2025},
 title = {Choosing an SBOM Generation Tool},
 url = {https://openssf.org/blog/2025/06/05/choosing-an-sbom-generation-tool/},
 lastaccessed = {2025-12-28},
 institution = {{Open Source Security Foundation}},
 year = {2025}
}

@misc{ncsc.sunburst.2021,
 author = {{National Cyber Security Centre}},
 date = {2021},
 title = {Dealing with the SolarWinds Orion compromise},
 url = {https://www.ncsc.gov.uk/guidance/dealing-with-the-solarwinds-orion-compromise},
 lastaccessed = {2026-03-11},
 year = {2021}
}

@misc{NIST.Def.SBOM.2026,
 author = {{National Institute of Standards and Technology (NIST)}},
 date = {2026},
 title = {SBOM},
 url = {https://csrc.nist.gov/glossary/term/sbom},
 lastaccessed = {2026-03-13},
 year = {2026}
}

@inproceedings{Nocera.2024,
 author = {Nocera, Sabato and {Di Penta}, Massimiliano and Francese, Rita and Romano, Simone and Scanniello, Giuseppe},
 title = {If it's not SBOM, then what? How Italian Practitioners Manage the Software Supply Chain},
 pages = {730--740},
 publisher = {IEEE},
 isbn = {979-8-3503-9568-6},
 booktitle = {2024 IEEE International Conference on Software Maintenance and Evolution (ICSME)},
 year = {2024},
 doi = {10.1109/ICSME58944.2024.00077},
 address = {Flagstaff, AZ, USA}
}

@misc{NTIA.2019b,
 author = {{National Telecommunications and Information Administration}},
 date = {2019},
 title = {Roles and Benefits for SBOM Across the Supply Chain: NTIA Multistakeholder Process on Software Component Transparency Use Cases and State of Practice Working Group},
 url = {https://www.ntia.gov/files/ntia/publications/ntia_sbom_use_cases_roles_benefits-nov2019.pdf},
 lastaccessed = {2025-08-27},
 institution = {{National Telecommunications and Information Administration}},
 year = {2019}
}

@misc{NTIA.2021b,
 author = {{National Telecommunications and Information Administration}},
 date = {2021},
 title = {The Minimum Elements For a Software Bill of Materials (SBOM)},
 url = {https://www.ntia.gov/report/2021/minimum-elements-software-bill-materials-sbom},
 lastaccessed = {2025-07-22},
 institution = {{National Telecommunications and Information Administration}},
 year = {2021}
}

@misc{NTIA.2021c,
 author = {{National Telecommunications and Information Administration}},
 date = {2021},
 title = {Survey of Existing SBOM Formats and Standards},
 url = {https://www.ntia.gov/sites/default/files/publications/sbom_formats_survey-version-2021_0.pdf},
 lastaccessed = {2025-08-20},
 institution = {{National Telecommunications and Information Administration}},
 year = {2021}
}

@misc{NTIA.2021e,
 author = {{National Telecommunications and Information Administration}},
 date = {2021},
 title = {Framing Software Component Transparency: Establishing a Common Software Bill of Materials (SBOM): Second Edition},
 url = {https://www.ntia.gov/files/ntia/publications/ntia_sbom_framing_2nd_edition_20211021.pdf},
 lastaccessed = {2025-11-02},
 institution = {{National Telecommunications and Information Administration}},
 year = {2021}
}

@misc{NTIA.Naming.2021,
 author = {{NTIA Multistakeholder Process on Software Component Transparency Framing Working Group}},
 date = {2021},
 title = {Software Identification  Challenges and Guidance},
 url = {https://www.ntia.gov/files/ntia/publications/ntia_sbom_software_identity-2021mar30.pdf},
 lastaccessed = {2026-03-09},
 year = {2021}
}

@misc{NVD,
 author = {{Information Technology Laboratory}},
 date = {2024},
 title = {National Vulnerability Database},
 url = {https://nvd.nist.gov/},
 lastaccessed = {2025-08-27},
 institution = {{National Institute of Standards and Technology (NIST)}},
 year = {2024}
}

@inproceedings{ODonoghue.2024,
 author = {O'Donoghue, Eric and Reinhold, Ann Marie and Izurieta, Clemente},
 title = {Assessing Security Risks of Software Supply Chains Using Software Bill of Materials},
 pages = {134--140},
 publisher = {IEEE},
 isbn = {979-8-3503-5157-6},
 booktitle = {2024 IEEE International Conference on Software Analysis, Evolution and Reengineering - Companion (SANER-C)},
 year = {2024},
 doi = {10.1109/SANER-C62648.2024.00023},
 address = {Rovaniemi, Finland}
}

@inproceedings{ODonoghue.2024b,
 author = {O'Donoghue, Eric and Boles, Brittany and Izurieta, Clemente and Reinhold, Ann Marie},
 title = {Impacts of Software Bill of Materials (SBOM) Generation on Vulnerability Detection},
 pages = {67--76},
 publisher = {ACM},
 isbn = {9798400712401},
 editor = {Torres-Arias, Santiago and de Carli, Lorenzo and Zhang, Yuchen},
 booktitle = {Proceedings of the 2024 Workshop on Software Supply Chain Offensive Research and Ecosystem Defenses},
 year = {2024},
 doi = {10.1145/3689944.3696164},
 address = {Salt Lake City UT USA}
}

@misc{OpenSBOM.Spdx,
 author = {Puerco},
 date = {2021},
 title = {SPDX Software Bill of Materials (SBOM) Generator},
 url = {https://github.com/opensbom-generator/spdx-sbom-generator},
 lastaccessed = {2025-07-21},
 year = {2021}
}

@misc{Oracle.2025,
 author = {Oracle},
 date = {2025},
 title = {Java Docs: Class System},
 url = {https://docs.oracle.com/en/java/javase/25/docs/api/java.base/java/lang/System.html},
 lastaccessed = {2025-10-02},
 year = {2025}
}

@misc{ORT.Adopters.2025,
 author = {{The ORT Project Authors}},
 date = {2025},
 title = {Adopters},
 url = {https://github.com/boschglobal/oss-review-toolkit/blob/main/ADOPTERS.md},
 lastaccessed = {2025-12-28},
 institution = {{The ORT Project Authors}},
 year = {2025}
}

@misc{ORT.Configuration.2025,
 author = {{The ORT Project Copyright Holders}},
 date = {2025},
 title = {OSS Review Toolkit: Repository Configuration (.ort.yml)},
 url = {https://oss-review-toolkit.org/ort/docs/configuration/ort-yml},
 lastaccessed = {2025-12-06},
 year = {2025}
}

@misc{ORT.Documentation.Poetry,
 author = {{The ORT Project Copyright Holders}},
 date = {2026},
 title = {ORT Documentation: Poetry},
 url = {https://oss-review-toolkit.org/ort/docs/plugins},
 lastaccessed = {2026-04-08},
 year = {2026}
}

@misc{Osquery.2026,
 author = {Osquery},
 date = {2026},
 title = {osquery},
 url = {https://github.com/osquery/osquery},
 lastaccessed = {2026-04-08},
 year = {2026}
}

@misc{OWASP.Blint,
 author = {{Open Web Application Security Project}},
 date = {2021},
 title = {blint},
 url = {https://github.com/owasp-dep-scan/blint},
 lastaccessed = {2025-13-11},
 year = {2021}
}

@misc{OWASP.Depscan,
 author = {{Open Web Application Security Project}},
 date = {2020},
 title = {OWASP dep-scan},
 url = {https://github.com/owasp-dep-scan/dep-scan},
 lastaccessed = {2025-07-21},
 editor = {{Open Web Application Security Project}},
 year = {2020}
}

@misc{OWASP.Top10.2025,
 author = {{Open Web Application Security Project}},
 date = {2025},
 title = {OWASP Top 10:2025},
 url = {https://owasp.org/Top10/2025/},
 lastaccessed = {2026-01-10},
 institution = {{Open Web Application Security Project}},
 year = {2025}
}

@misc{Ozkan.2024,
 author = {Ozkan, Can and Zou, Xinhai and Singelee, Dave},
 date = {2024},
 title = {Supply Chain Insecurity: The Lack of Integrity Protection in SBOM Solutions},
 publisher = {arXiv},
 doi = {10.48550/arXiv.2412.05138},
 year = {2024}
}

@misc{Paloalto.2015,
 author = {Xiao, Claud},
 date = {2015},
 title = {Novel Malware XcodeGhost Modifies Xcode, Infects Apple iOS Apps and Hits App Store},
 url = {https://unit42.paloaltonetworks.com/novel-malware-xcodeghost-modifies-xcode-infects-apple-ios-apps-and-hits-app-store/},
 lastaccessed = {2026-03-10},
 institution = {{Paloalto Networks}},
 year = {2015}
}

@misc{PHP.Composer,
 author = {{The PHP Documentation Group}},
 date = {2026},
 title = {Introduction to Composer},
 url = {https://www.php.net/manual/en/install.composer.intro.php},
 lastaccessed = {2026-01-11},
 year = {2026}
}

@misc{PHP.eval,
 author = {{The PHP Documentation Group}},
 date = {2025},
 title = {PHP Language Reference: eval},
 url = {https://www.php.net/manual/en/function.eval.php},
 lastaccessed = {2025-10-03},
 year = {2025}
}

@misc{pip.2025,
 author = {{The pip developers}},
 date = {2025},
 title = {pip documentation: User Guide},
 url = {https://pip.pypa.io/en/stable/user_guide/},
 lastaccessed = {2025-10-01},
 year = {2025}
}

@misc{pip.requirements.txt,
 author = {{The pip developers}},
 date = {2026},
 title = {pip documentation: Requirements File Format},
 url = {https://pip.pypa.io/en/stable/reference/requirements-file-format/},
 lastaccessed = {2026-01-11},
 year = {2026}
}

@misc{Poetry.2025,
 author = {Poetry},
 date = {2025},
 title = {Poetry: Dependency specification},
 url = {https://python-poetry.org/docs/main/managing-dependencies/},
 lastaccessed = {2025-10-02},
 year = {2025}
}

@misc{Prabhu.2024,
 author = {Prabhu},
 date = {2024},
 title = {[Security] Code execution risk when running cdxgen against untrusted repos {\#}1328},
 url = {https://github.com/cdxgen/cdxgen/issues/1328},
 lastaccessed = {2026-04-14},
 year = {2024}
}

@misc{PyPi.Stats,
 author = {{Python Software Foundation}},
 date = {2026},
 title = {PyPI Stats: Analytics for PyPI packages},
 url = {https://pypistats.org/},
 lastaccessed = {2026-03-10},
 year = {2026}
}

@misc{Python.2025b,
 author = {{Python Software Foundation}},
 date = {2025},
 title = {The Python Standard Library: Build-in Functions},
 url = {https://docs.python.org/3/library/functions.html},
 lastaccessed = {2025-10-03},
 year = {2025}
}

@misc{PythonPackagingAuthority.2025,
 author = {{Python Packaging Authority}},
 date = {2025},
 title = {pip},
 url = {https://pip.pypa.io/en/stable/},
 lastaccessed = {2025-11-12},
 year = {2025}
}

@inproceedings{Rabbi.2024,
 author = {Rabbi, Md Fazle and Champa, Arifa Islam and Nachuma, Costain and Zibran, Minhaz Fahim},
 title = {SBOM Generation Tools Under Microscope: A Focus on The npm Ecosystem},
 pages = {1233--1241},
 publisher = {ACM},
 isbn = {9798400702433},
 editor = {Hong, Jiman and Park, Juw Won and Przyby{\l}ek, Adam},
 booktitle = {Proceedings of the 39th ACM/SIGAPP Symposium on Applied Computing (SAC)},
 year = {2024},
 doi = {10.1145/3605098.3635927},
 address = {Avila Spain}
}

@misc{Rust.manifest.2026,
 author = {{The Rust Programming Language}},
 date = {2026},
 title = {The Cargo Book: The Manifest Format},
 url = {https://doc.rust-lang.org/cargo/reference/manifest.html},
 lastaccessed = {2026-04-08},
 year = {2026}
}

@misc{SBOM4Python,
 author = {Harrison, Anthony},
 date = {2022},
 title = {SBOM4Python},
 url = {https://pypi.org/project/sbom4python/},
 lastaccessed = {2025-07-21},
 year = {2022}
}

@misc{SBOM4Rust,
 author = {Harrison, Anthony},
 date = {2022},
 title = {SBOM4Rust},
 url = {https://github.com/anthonyharrison/sbom4rust},
 lastaccessed = {2025-07-21},
 year = {2022}
}

@misc{Scancode,
 author = {AboutCode},
 date = {2026},
 title = {scancode-toolkit},
 url = {https://github.com/aboutcode-org/scancode-toolkit},
 lastaccessed = {2026-03-22},
 year = {2026}
}

@misc{ScribeSecurity.Scribe,
 author = {{Scribe Security}},
 date = {2025},
 title = {Scribe},
 url = {https://scribesecurity.com/de/scribe-platform/},
 lastaccessed = {2025-07-21},
 institution = {{Scribe Security}},
 year = {2025}
}

@misc{Sham.2025,
 author = {Sham, Swaroop},
 date = {2025},
 title = {The Top 11 Open-Source SBOM tools},
 url = {https://www.wiz.io/academy/application-security/top-open-source-sbom-tools},
 lastaccessed = {2025-12-28},
 institution = {{Wiz Inc.}},
 year = {2025}
}

@article{Shu.2025,
 author = {Shu, Congyan and Chen, Wentao and Fan, Guisheng and Yu, Huiqun and Huang, Zijie and Liang, Yuguo},
 year = {2025},
 title = {Tool or Toy: Are SCA tools ready for challenging scenarios?},
 pages = {104624},
 volume = {158},
 issn = {01674048},
 journal = {Computers {\&} Security},
 doi = {10.1016/j.cose.2025.104624}
}

@misc{Snyk,
 author = {{Snyk Limited}},
 date = {2025},
 title = {Snyk},
 url = {https://snyk.io/},
 lastaccessed = {2025-07-21},
 institution = {{Snyk Limited}},
 year = {2025}
}

@misc{Sonatype.2023,
 author = {Sonatype},
 date = {2023},
 title = {Comparing SBOM standards: SPDX vs. CycloneDX},
 url = {https://www.sonatype.com/blog/comparing-sbom-standards-spdx-vs.-cyclonedx-vs.-swid},
 lastaccessed = {2025-08-20},
 institution = {Sonatype},
 year = {2023}
}

@misc{Sonatype.2026,
 author = {Sonatype},
 date = {2026},
 title = {State of the Software Supply Chain: 2026},
 url = {https://www.sonatype.com/state-of-the-software-supply-chain/2026/software-infrastructure-growth},
 lastaccessed = {2026-03-10},
 institution = {Sonatype},
 year = {2026}
}

@misc{SPDX,
 author = {{The Linux Foundation}},
 date = {2023},
 title = {SPDX Specifications},
 url = {https://spdx.dev/use/specifications/},
 lastaccessed = {2025-08-20},
 institution = {{The Linux Foundation}},
 year = {2023}
}

@misc{SPDX.NTIA.ComplianceChecker.2022,
 author = {{The Linux Foundation}},
 date = {2022},
 title = {SPDX NTIA Conformance Checker},
 url = {https://github.com/spdx/ntia-conformance-checker},
 lastaccessed = {2025-10-25},
 year = {2022}
}

@misc{SPDX.package.2024,
 author = {{The Linux Foundation}},
 date = {2024},
 title = {SPDX Specification: Package},
 url = {https://spdx.github.io/spdx-spec/v3.0.1/model/Software/Classes/Package/},
 lastaccessed = {2025-12-12},
 year = {2024}
}

@misc{SPDX.Python.Tools.2026,
 author = {SPDX},
 date = {2026},
 title = {tools-python},
 url = {https://github.com/spdx/tools-python},
 lastaccessed = {2026-03-09},
 year = {2026}
}

@misc{SPDX.Python.Tools.Documentation.2026,
 author = {SPDX},
 date = {2026},
 title = {tools-python: Code architecture documentation},
 url = {https://github.com/spdx/tools-python/blob/main/DOCUMENTATION.md},
 lastaccessed = {2026-03-09},
 year = {2026}
}

@misc{SPDX.Toolings.2016,
 author = {{The Linux Foundation}},
 date = {2016},
 title = {SPDX tools-python},
 url = {https://github.com/spdx/tools-python},
 lastaccessed = {2025-10-09},
 year = {2016}
}

@misc{Springett.2025,
 author = {Springett, Steve},
 date = {2025},
 title = {Advisory on Software Bill of Materials and Real-time Vulnerability Monitoring for Open-Source Software and Third-Party Dependencies},
 url = {https://owasp.org/blog/2025/02/24/advisory-on-implementation-of-software-bill-of-materials-for-vulnerability-management},
 lastaccessed = {2025-08-20},
 institution = {{Open Web Application Security Project}},
 year = {2025}
}

@inproceedings{Stalnaker.2024,
 author = {Stalnaker, Trevor Wayne and Wintersgill, Nathan and Chaparro, Oscar and {Di Penta}, Massimiliano and German, Daniel M. and Poshyvanyk, Denys},
 title = {BOMs Away! Inside the Minds of Stakeholders: A Comprehensive Study of Bills of Materials for Software Systems},
 pages = {1--13},
 publisher = {ACM},
 isbn = {9798400702174},
 editor = {Roychoudhury, Abhik and Paiva, Ana and Abreu, Rui and Storey, Margaret},
 booktitle = {Proceedings of the IEEE/ACM 46th International Conference on Software Engineering (ICSE)},
 year = {2024},
 doi = {10.1145/3597503.3623347},
 address = {Lisbon Portugal}
}

@misc{Svensson.covenant,
 author = {Svensson, Patrik},
 date = {2022},
 title = {Covenant},
 url = {https://github.com/patriksvensson/covenant},
 lastaccessed = {2025-07-21},
 year = {2022}
}

@misc{Syft,
 author = {Anchore},
 date = {2020},
 title = {Syft},
 url = {https://github.com/anchore/syft},
 lastaccessed = {2025-07-21},
 institution = {Anchore},
 year = {2020}
}

@inproceedings{Szekeres.2013,
 author = {Szekeres, L. and Payer, M. and Wei, Tao and Song, Dawn},
 title = {SoK: Eternal War in Memory},
 pages = {48--62},
 publisher = {IEEE},
 isbn = {978-0-7695-4977-4},
 booktitle = {2013 IEEE Symposium on Security and Privacy},
 year = {2013},
 doi = {10.1109/SP.2013.13},
 address = {Berkeley, CA}
}

@misc{TernTools.Tern,
 author = {{Tern Tools}},
 date = {2017},
 title = {Tern},
 url = {https://github.com/tern-tools/tern},
 lastaccessed = {2025-07-21},
 institution = {{Tern Tools}},
 year = {2017}
}

@misc{Thomas.2025,
 author = {Thomas, Romain},
 date = {2025},
 title = {LIEF},
 url = {https://github.com/lief-project/LIEF},
 lastaccessed = {2025-12-15},
 year = {2025}
}

@article{TorresArias.2023,
 author = {Torres-Arias, Santiago and Geer, Dan and Meyers, John Speed},
 year = {2023},
 title = {A Viewpoint on Knowing Software: Bill of Materials Quality When You See It},
 pages = {50--54},
 volume = {21},
 number = {6},
 issn = {1540-7993},
 journal = {IEEE Security {\&} Privacy},
 doi = {10.1109/MSEC.2023.3315887}
}

@misc{TR-03183-2.2024,
 author = {{Federal Office for Information Security}},
 date = {2024},
 title = {Technical Guideline TR-03183:  Cyber Resilience Requirements for Manufacturers and Products: Part 2: Software Bill of Materials (SBOM)},
 url = {https://www.bsi.bund.de/SharedDocs/Downloads/EN/BSI/Publications/TechGuidelines/TR03183/BSI-TR-03183-2-2_0_0.pdf},
 lastaccessed = {2025-03-17},
 edition = {2.0.0},
 institution = {{Federal Office for Information Security}},
 year = {2024}
}

@inproceedings{Tran.2024,
 author = {Tran, Nguyen Khoi and Pallewatta, Samodha and Babar, Muhammad Ali},
 title = {An Empirically Grounded Reference Architecture for Software Supply Chain Metadata Management},
 pages = {38--47},
 publisher = {ACM},
 isbn = {9798400717017},
 booktitle = {Proceedings of the 28th International Conference on Evaluation and Assessment in Software Engineering (EASE)},
 year = {2024},
 doi = {10.1145/3661167.3661212},
 address = {Salerno Italy}
}

@misc{Trivy,
 author = {Aquasecurity},
 date = {2019},
 title = {Trivy},
 url = {https://github.com/aquasecurity/trivy},
 lastaccessed = {2025-07-21},
 institution = {Aquasecurity},
 year = {2019}
}

@misc{Trivy.Languages,
 author = {Aquasecurity},
 date = {2025},
 title = {Trivy: Programming Language},
 url = {https://trivy.dev/latest/docs/coverage/language/},
 lastaccessed = {2025-10-24},
 year = {2025}
}

@misc{Umbelino.2022,
 author = {Umbelino, Pedro},
 date = {2022},
 title = {Log4j Vulnerability (Log4Shell): Ongoing Challenges in Remediation},
 url = {https://www.bitsight.com/blog/log4j-vulnerability-log4shell-ongoing-challenges-remediation},
 lastaccessed = {2025-08-20},
 institution = {Bitsight},
 year = {2022}
}

@misc{UpwindSecurity.2025,
 author = {Burgin, Joshua},
 date = {2025},
 title = {The Top 6 Open-Source SBOM Tools},
 url = {https://www.upwind.io/glossary/the-top-6-open-source-sbom-tools},
 lastaccessed = {2025-12-28},
 institution = {{Upwind Security}},
 year = {2025}
}

@misc{Vaas.2022,
 author = {Vaas, Lisa},
 date = {2022},
 title = {One year after Log4Shell, firms still struggle to hunt down Log4j},
 url = {https://www.contrastsecurity.com/security-influencers/one-year-after-log4shell-firms-still-struggle-to-hunt-down-log4j},
 lastaccessed = {2025-08-20},
 institution = {{Contrast Security}},
 year = {2022}
}

@article{Wang.2026b,
 author = {Wang, Chengjie and Wu, Jingzheng and Lyu, Hao and Ling, Xiang and Luo, Tianyue and Wu, Yanjun and Zhao, Chen},
 year = {2026},
 title = {A Large Scale Empirical Analysis on the Adherence Gap between Standards and Tools in SBOM},
 url = {https://dl.acm.org/doi/epdf/10.1145/3788692},
 urldate = {2026-03-22},
 issn = {1049-331X},
 journal = {ACM Transactions on Software Engineering and Methodology},
 doi = {10.1145/3788692}
}

@inproceedings{Xiao.2025,
 author = {Xiao, Yue and Kirat, Dhilung and Schales, Douglas Lee and Jang, Jiyong and Xing, Luyi and Liao, Xiaojing},
 title = {JBomAudit: Assessing the Landscape, Compliance, and Security Implications of Java SBOMs},
 pages = {1--20},
 publisher = {{Internet Society}},
 isbn = {979-8-9894372-8-3},
 editor = {P{\"o}pper, Christina and Okhravi, Hamed},
 booktitle = {Proceedings 2025 Network and Distributed System Security Symposium},
 year = {2025},
 doi = {10.14722/ndss.2025.240322},
 address = {San Diego, CA, USA}
}

@inproceedings{Yu.2024,
 author = {Yu, Sheng and Song, Wei and Hu, Xunchao and Yin, Heng},
 title = {On the Correctness of Metadata-Based SBOM Generation: A Differential Analysis Approach},
 pages = {29--36},
 publisher = {IEEE},
 isbn = {979-8-3503-4105-8},
 booktitle = {2024 54th Annual IEEE/IFIP International Conference on Dependable Systems and Networks (DSN)},
 year = {2024},
 doi = {10.1109/DSN58291.2024.00018},
 address = {Brisbane, Australia}
}

@article{Zahan.2023,
 author = {Zahan, Nusrat and Lin, Elizabeth and Tamanna, Mahzabin and Enck, William and Williams, Laurie},
 year = {2023},
 title = {Software Bills of Materials Are Required. Are We There Yet?},
 pages = {82--88},
 volume = {21},
 number = {2},
 issn = {1540-7993},
 journal = {IEEE Security {\&} Privacy},
 doi = {10.1109/MSEC.2023.3237100},
}

\appendices

\section{Reproducibility}
We publish our assessment to enable reproducibility and to provide a framework that can be extended and used to continuously evaluate the capabilities of SBOM tools, thereby improving their further development.

To enable easy reproducibility of our results, we provide (a) a Docker container that contains the environment used, (b) the scripts used, and (c) the tool outputs and logs.

In addition to this replication package, we will publish the scripts separately on GitHub when the paper is published to enable further development of the assessment.

To reproduce our results, please proceed as follows:
\begin{enumerate}[leftmargin=*, label=\arabic*.]
    \item Download the Docker image from: Pending Publication

    \item Extract the archive.

    \item Load the Docker image:
\begin{lstlisting}[language=bash]
docker load -i sbom-benchmarking-latest.tar.gz
\end{lstlisting}

    \item Create an output directory:
\begin{lstlisting}[language=bash]
mkdir -p output
\end{lstlisting}

    \item Execute the tests:
\begin{lstlisting}[language=bash]
docker run --rm -it \
  -v $(pwd)/output:/app/output \
  -w /app \
  sbom-benchmarking:latest \
  python /app/generate_sboms.py
\end{lstlisting}

    \item Compare the results listed in \texttt{latex\_component\_detection.tex}
    with Table~\ref{tab:latex_component_detection}.
\end{enumerate}

\section{Selected Programming Languages}\label{rel:SelectedProgrammingLanguages}
We selected five popular and common programming languages to evaluate our selection of SBOM generators against.
Our selection is based on popularity, relevance, and technology:
PHP is still the dominant server-side web language~\cite{web_technologies_overview}, whereas Java is used by a majority of companies~\cite{java_usage}.
Python is highly popular for scripting and a common entry-level programming language~\cite{python_usage}, whereas Go and Rust are modern, security-by-design languages that produce binaries~\cite{go_documentation}, unlike the other languages that require an interpreter or a VM.

\emph{Python} is an interpreted, high-level, general-purpose programming language that has seen a sharp rise in popularity over the last few years due to the availability of powerful machine learning libraries.
Python is a mature programming language that was first released in 1991~\cite{link:pythonHistory}.
We developed our tests against Python 3, the most recent version of Python.
Based on a survey of developers, \empirical{\num{54}\%} of programmers use Python regularly~\cite{JetBrains.2023}.

\emph{Java} is a compiled, general-purpose, and high-level programming language that is commonly deployed for business applications~\cite{java_usage}.
Its source code is compiled into bytecode that is then run in the Java Virtual Machine~\cite{java_jvm}.
It is a mature language first released in 1996~\cite{java_first_release}.
Based on a survey of developers, \empirical{\num{49}\%} of programmers use Java regularly~\cite{JetBrains.2023}.

\emph{C} is a compiled, low-level, general-purpose programming language developed.
It remains widely adopted, used by 19\% of developers~\cite{JetBrains.2023}, despite being developed already in the 1970s.
Because C provides low-level hardware access, it enables highly optimized execution, making it a commonly chosen language for performance-critical applications, even outperforming languages like Rust in most cases~\cite{Li.2024b}.
At the same time, there are few built-in automated security checks, meaning that buffer overflows remain a challenge~\cite{Szekeres.2013}.

\emph{Go} is a compiled, high-level, general-purpose programming language initially designed by Google as an answer to recurring programming challenges, including multicore processors, networked systems, massive computation clusters, and the web programming model~\cite{link:golang}.
It is one of the youngest of our programming languages and was first released in 2009~\cite{go.faq}.
Based on a survey of developers, \empirical{\num{17}\%} of programmers use Go regularly~\cite{JetBrains.2023}.

\emph{Rust} is similar to Go, a compiled, high-level, general-purpose programming language that focuses on security-by-design, ensuring that code written in it cannot exhibit common vulnerability-related problems such as faulty memory management.
It was released in 2012 and is thus the youngest of our programming languages, but still relevant with \empirical{\num{10}\%} of programmers that use Rust regularly~\cite{JetBrains.2023} and large projects such as Firefox~\cite{Mozilla.2017} and the Linux Kernel~\cite{corbet.2025} incorporating features implemented in Rust.

\emph{PHP} is an interpreted, primarily server-side used, programming language that runs over 
\empirical{\num{70}\%} of all websites~\cite{web_technologies_overview} and, thus, forms a backbone of the current Internet.
It was released in 1995~\cite{php.history}, and \empirical{\num{18}\%} of developers report using it regularly~\cite{JetBrains.2023}.

\newpage

\section{SBOM Types}\label{apx:sbomTypes}
\begin{table}[!h]
\rowcolors{2}{white}{gray!15}
\begin{tabularx}{\columnwidth}{lp{6.3cm}}
\small
\textbf{SBOM Type} & \textbf{Created ...} \\ \midrule
Design    & from the planned set of included components of a software project, without the components actually already needing to exist. \\
Source    & from the development environment and source files.\\
Build     & during the build process using source files, dependency information, build process data, and other related SBOMs.\\
Analyzed  & based on the build outcome using the created artifacts such as executables, packages, containers or VM images.\\
Deployed  & based on the system software is deployed on possibly in combination with other SBOMs.\\
Runtime   & based on the execution of the software, capturing executed components, external calls, and dynamically loaded components.\\
\end{tabularx}
\caption{Overview of the six SBOM types provided by the CISA and BSI~\cite{CISA.2023b, TR-03183-2.2024}}\label{tbl:sbomtypes}
\end{table}

\section{Capability Overview}
We provide a lists the approaches and files used by the evaluated SBOM generators to identify components in software projects in Table~\ref{tbl:analyzedFiles}.

\section{Related Work Table}
We provide an overview of the academic literature that evaluated the quality of SBOMs in Table~\ref{tab:related_work}.

\section{Toolings Used in Literature}
We provide an overview of the tools evaluated in the broader academic literature in Table~\ref{tab:sbom-tools}.

\begin{table*}[!ht]
    \small
    \begin{tabularx}{\textwidth}{lX}
    \textbf{Approach} & \textbf{Files}              \\ \midrule
    \multicolumn{2}{c}{\textbf{Python}} \\ \midrule
    .py             & *.py (source code)\filescdxgen \\ \midrule               
    wheel/egg       & *.whl\filescdxgen\filessyft\filestrivy, *.egg-info\filescdxgen\filessyft\filestrivy, *.egg\filestrivy, setup.py\filescdxgen\filessyft\filesort\filessbomtool                  \\ \midrule
    Pip             & requirements.txt\filescdxgen\filessyft\filestrivy\filesort\filessbomtool         \\ \midrule
    conda           & environment.yml\filescdxgen\filessyft\filestrivy\filesort\filessbomtool \\ \midrule
    Pixi            & pixi.toml\filescdxgen, pixi.lock\filescdxgen \\ \midrule
    Pipenv          & Pipfile.lock\filescdxgen\filessyft\filestrivy\filesort                             \\ \midrule
    Poetry          & poetry.lock\filescdxgen\filessyft\filestrivy\filesort\filessbomtool, pyproject.toml\filescdxgen\filesort         \\ \midrule
    pdm             & pdm.lock\filescdxgen\filessyft                           \\ \midrule
    Hatch           & pyproject.toml\filescdxgen                           \\ \midrule
    uv              & uv.lock\filescdxgen\filessyft\filestrivy         \\ \midrule 
    venv            & virtual environment\filescdxgen\filessyft \\ \midrule  
    \multicolumn{2}{c}{\textbf{Java}} \\
    \midrule
    
    JVM archives    & jar\filescdxgen\filessyft\filestrivy, war\filescdxgen\filessyft\filestrivy, ear\filessyft\filestrivy, par\filessyft\filestrivy, sar\filessyft, nar\filessyft, jpi\filessyft, hpi\filescdxgen\filessyft, kar\filessyft, lpkg\filessyft rar\filessyft \\ \midrule
    Maven           & pom.xml\filescdxgen\filessyft\filestrivy\filesort\filessbomtool, maven-index\filescdxgen, maven-cache\filescdxgen, maven.xml\filescdxgen                     \\ \midrule
    Gradle          & build.gradle\filescdxgen\filesort, build.gradle.kts\filescdxgen\filesort, gradle.lock\filessyft\filestrivy, gradle.lockfile\filessbomtool, settings.gradle\filescdxgen\filesort, settings.gradle.kts\filescdxgen\filesort, gradle-index\filescdxgen, gradle-cache\filescdxgen                \\ \midrule
    Quarkus         & pom.xml\filescdxgen, build.gradle\filescdxgen, build.mill\filescdxgen                            \\ \midrule
    Bazel           & WORKSPACE.bazel\filescdxgen, MODULE.bazel\filescdxgen \\ \midrule
    Mill            & build.mill\filescdxgen                            \\ \midrule
    SBT             & build.sbt\filescdxgen\filesort, *.sbt.lock\filestrivy, build.scala\filescdxgen\filesort, sbt-index\filescdxgen, sbt-cache\filescdxgen                   \\ \midrule
    
    \multicolumn{2}{c}{\textbf{Go}} \\
    \midrule
    
    Go mod          & go.mod\filescdxgen\filessyft\filestrivy\filesort\filessbomtool        \\ \midrule
    Go Lock         & Gopkg.lock\filescdxgen                        \\ \midrule
    Go Sum          & go.sum\filescdxgen\filessbomtool \\ \midrule
    Go binary       & (binary file)\filescdxgen\filessyft\filestrivy               \\ \midrule
    
    \multicolumn{2}{c}{\textbf{PHP}} \\
    \midrule
    
    Composer        & composer.json\filesort, composer.lock\filescdxgen\filessyft\filestrivy\filesort \\ \midrule
    PHP Installed   & installed.json\filessyft\filestrivy \\ \midrule
    PECL            & *.reg\filessyft \\ \midrule
    PEAR            & *.reg\filessyft \\ \midrule

    \multicolumn{2}{c}{\textbf{Rust}} \\
    \midrule
    
    Cargo           & Cargo.lock\filescdxgen\filessyft\filestrivy\filessbomtool, Cargo.toml\filescdxgen\filessyft\filestrivy\filessbomtool \\ \midrule
    Rust binary     & (binary file)\filescdxgen\filessyft\filestrivy \\ \midrule
    
    \multicolumn{2}{c}{\textbf{C/C++}} \\
    \midrule
    
    Conan           & conan.lock\filescdxgen\filessyft\filestrivy\filessbomtool, conanfile.txt\filescdxgen\filessyft, conaninfo.txt\filessyft \\ \midrule
    CMake           & CMakeLists.txt\filescdxgen, *.cmake\filescdxgen \\ \midrule
    Meson           & meson.build\filescdxgen \\ \midrule
    vcpkg           & vcpkg.json\filescdxgen\filessbomtool, vcpkg.spdx.json\filessbomtool \\ \midrule
    Source Files    & *.c (source code)\filescdxgen, *.h (headers)\filescdxgen \\ \midrule
    C/C++ binary    & (binary file)\filescdxgen\filessyft \\ \bottomrule
    
    \end{tabularx}
    \caption{Overview of the different approaches to identify components. Dots indicate the tooling support: \cdxgen\textcolor{cbcdxgen}{\Large\textbullet}, \syft\textcolor{cbsyft}{\Large\textbullet}, \trivy\textcolor{cbtrivy}{\Large\textbullet}, \ort\textcolor{cbort}{\Large\textbullet}, \sbomtool\textcolor{cbsbomtool}{\Large\textbullet}. Some tools analyze the output of project management tools rather than the files directly.}\label{tbl:analyzedFiles}
\end{table*}
\begin{table*}[!ht]
\centering
\resizebox{\textwidth}{!}{%
\small
\rowcolors{2}{gray!15}{white}
\begin{tabularx}{\textwidth}{
l|
>{\hsize=1.3\hsize\linewidth=\hsize}X|
>{\hsize=0.9\hsize\linewidth=\hsize}X|
>{\hsize=0.9\hsize\linewidth=\hsize}X|
>{\hsize=0.9\hsize\linewidth=\hsize}X
}
        \textbf{Paper} & \textbf{Summary} & \textbf{Scope} & \textbf{Venue} & \textbf{Tools}\\

        \hline
        \rowcolor{white}
        \multicolumn{5}{c}{\textbf{Synthetic test scripts}}\\
        \hline

        Cofano et al.~\cite{Cofano.2024} & Created Python scripts integrating managed dependencies and evaluated the resulting SBOMs to identify blind spots & 12 synthetic Python projects & TrustCom 2024 & trivy, syft, cdxgen, ORT\\ 

        Garcia et al.~\cite{Garcia.2025} & Describes the SBOM Tool landscape and evaluated 5 tools against 3 synthetic tests for managed components & 5 tools & SVM 2025 & trivy, syft, cdxgen, ORT, SPDX SBOM Generator\\

        \hline
        \rowcolor{white}
        \multicolumn{5}{c}{\textbf{Analyze Source-Code/Binaries}}\\
        \hline
        
        Xiao et al.~\cite{Xiao.2025} & Analyzed JAR files and compared the results to existing associated SBOMs & 25.8K Java SBOMs & NDSS 2025 & n/a\\

        Wang et al.~\cite{Wang.2026b} & Evaluated the compliance with frameworks and the consistency between tools for real-world projects. Additionally, the ground truth was generated from Python code to evaluate completeness. & 3.2K Github Projects & TOSEM & cdxgen, gh-sbom, ort, scancode, syft \\

        Kong et al.~\cite{Kong.2025} & Developed a tool to extract components from source code and compared it to the output of 4 different SBOM generators & 30 projects & Scientific reports & cdxgen, sbom-tool, OpenRewrite, build-info-go \\

        \hline
        \rowcolor{white}
        \multicolumn{5}{c}{\textbf{Use project management tools to generate a component list}}\\
        \hline

        Balliu et al.~\cite{Balliu.2023} & Used the maven-dependencyplugin to produce a ground truth for comparison & 26 Java Maven Projects & IEEE Security \& Privacy Journal & Build-Info-Go, cdxgen, Cdx-maven-plugin, Deepscan, jbom, OpenRewrite\\

        Rabbi et al.~\cite{Rabbi.2024} & Used the npm list command to produce a ground truth for comparison & 50 JavaScript npm projects & SAC 2024 & ORT, cnn, syft, cdxgen\\
        
        \hline
        \rowcolor{white}
        \multicolumn{5}{c}{\textbf{Compare against NTIA minimum elements}}\\
        \hline
        Torres-Arias et al.~\cite{TorresArias.2023} & Evaluated the existing fields of SBOMs generated with 4 tools by using the NTIA Conformance Checker and the SBOM Score Card & 1000 Docker Hub Container Images & IEEE Security \& Privacy Journal & syft, bom, tern, trivy\\
        
        Halbritter and Merli~\cite{Halbritter.2024} & Verified the NTIA compliance for 4 self-written projects and evaluated if all managed components were included & Python, C, Rust, Typescript & ARES 2024 & Various Tools\\
        
        Muneza et al.~\cite{ManziMuneza.2025} & Compared against NTIA minimum elements & 2,2K Container Images & SANER-C 2025 & Trivy, Syft\\

        \hline
        \rowcolor{white}
        \multicolumn{5}{c}{\textbf{Differential Analysis}}\\
        \hline

        Yu et al.~\cite{Yu.2024} & Compared the results from different tools (Differential Analysis) & 7.8K SBOMs in Python, Ruby, PHP, Java, Swift, C\#, Rust, Golang, JavaScript & DSN 2024 & trivy, syft, Microsoft SBOM Tool, GitHub Dependency Graph\\

        \hline
        \rowcolor{white}
        \multicolumn{5}{c}{\textbf{Bidirectional component cross-check}}\\
        \hline

        Bifolco et al.~\cite{Bifolco.2024} & Evaluated if dependencies list a specific project (backward) and if dependents are listed as dependencies (forward). & 635 GitHub Dependency Graphs of Java and Python Projects & EASE 2024 & Github Dependency Graph\\

        




\end{tabularx}%
}
\caption{Overview of academic literature evaluating SBOM quality grouped by the main evaluation approach.}
\label{tab:related_work}
\end{table*}
\begin{table*}[ht!]
\centering
\resizebox{\textwidth}{!}{%
\begin{tabular}{|l|c|c|c|c|c|c|c|c|c|c|c|c|c|c|c|c|c|c|c|c|c|c|c|c|c|c|c|c|}
\hline
\textbf{Paper/Tool} & \rotatebox{90}{Syft~\cite{Syft}} & \rotatebox{90}{CycloneDX Generator (cdxgen)~\cite{cdxgen}} & \rotatebox{90}{Trivy~\cite{Trivy}} & \rotatebox{90}{Microsoft SBOM Tool~\cite{MS.SBOMTOOL}} & \rotatebox{90}{GitHub Dependency Graph~\cite{GitHubDependencyGraph}} & \rotatebox{90}{Build-Info-Go~\cite{JFrog}} & \rotatebox{90}{CycloneDX-Maven-Plugin~\cite{CDX.MavenPlugin}} & \rotatebox{90}{CycloneDX-Conan-Extension~\cite{CDX.Conan}} & \rotatebox{90}{CycloneDX .Net~\cite{CDX.DotNet}} & \rotatebox{90}{CycloneDx for PHP~\cite{CDX.PHP}} & \rotatebox{90}{CycloneDX-Npm~\cite{CDX.npm}} & \rotatebox{90}{CycloneDX-Python~\cite{CDX.Python}} & \rotatebox{90}{Owasp dep-scan~\cite{OWASP.Depscan}} & \rotatebox{90}{jbom~\cite{ContrastSecurity.jbom}} & \rotatebox{90}{OpenRewrite~\cite{Moderne.Rewrite}} & \rotatebox{90}{OSS Review Toolkit (ORT)~\cite{LinuxFoundation.ORT}} & \rotatebox{90}{GH-sbom~\cite{AdvancedSecurity.GH-SBOM}} & \rotatebox{90}{OpenSBOM's Spdx~\cite{OpenSBOM.Spdx}} & \rotatebox{90}{kubernetes-sigs~\cite{Kubernetes.BOM}} & \rotatebox{90}{Tern~\cite{TernTools.Tern}} & \rotatebox{90}{SBOM4Python~\cite{SBOM4Python}} & \rotatebox{90}{SBOM4Rust~\cite{SBOM4Rust}} & \rotatebox{90}{Covenant~\cite{Svensson.covenant}} & \rotatebox{90}{Scribe~\cite{ScribeSecurity.Scribe}} & \rotatebox{90}{Chainguard Enforce Platform~\cite{Chainguard}} & \rotatebox{90}{Snyk~\cite{Snyk}} & \rotatebox{90}{FOSSA~\cite{FOSSA}} & \rotatebox{90}{scancode~\cite{Scancode}}\\
\hline
Balliu et al.~\cite{Balliu.2023}                            &               & \checkmark    &               &               &               & \checkmark        & \checkmark & & & & & & \checkmark & \checkmark & \checkmark & & & & & & & & & & & & & \\
\hline
Benedetti et al.~\cite{Benedetti.2025b}                      & \checkmark    & \checkmark    & \checkmark    &               &               &                   & & & & & & & & & & \checkmark & & & & & & & & & & & & \\
\hline
Cofano et al.~\cite{Cofano.2024}                            & \checkmark    & \checkmark    & \checkmark    &               &               &                   & & & & & & & & & & \checkmark & & & & & & & & & & & & \\
\hline
Crawford et al.~\cite{Crawford.2023} (Extended Abstract)    & \checkmark    &               &               &               &               &                   & & & & & & & & & & & & & & & & & & & & & & \\
\hline
Dalia et al.~\cite{Dalia.2024}                              & \checkmark    &               &               & \checkmark    &               &                   & \checkmark & & & & & & & & & & & \checkmark & \checkmark & \checkmark & & & & & & & & \\
\hline
Garcia et al.~\cite{Garcia.2025}                               & \checkmark    &   \checkmark            &   \checkmark            &    &               &                   &  & & & & & & & & & \checkmark & & \checkmark &  &  & & & & & & & & \\
\hline
Halbritter and Merli~\cite{Halbritter.2024}                 & \checkmark    & \checkmark    &               &               &               &                   & & \checkmark & & & \checkmark & \checkmark & & & & & & & & & \checkmark & \checkmark & \checkmark & & & & & \\
\hline
Kagzmandere and Arslan~\cite{Kagzmandere.2024}              &               &               &               & \checkmark    &               &                   & & & & & & & & & & & & & & & & & & & & & & \\
\hline
Kawaguchi and Hart~\cite{Kawaguchi.2024}                    & \checkmark    &               &               &               &               &                   & & & & & & & & & & & & & & & & & & & & & & \\
\hline
Mens and Decan~\cite{Mens.2024}                             & \checkmark    &               & \checkmark    & \checkmark    & \checkmark    &                   & & & & & & & & & & & & & & & & & & & & & & \\
\hline
O'Donoghue et al.~\cite{ODonoghue.2024b}                    & \checkmark    &               & \checkmark    &               &               &                             & & & & & & & & & & & & & & & & & & & & & & \\
\hline
O'Donoghue et al.~\cite{ODonoghue.2024}                     &               &               & \checkmark    &               &               &                             & & & & & & & & & & & & & & & & & & & & & & \\
\hline
Ozkan et al.~\cite{Ozkan.2024}                              & \checkmark    &               &               &               &               &                             & \checkmark & \checkmark & \checkmark & \checkmark & \checkmark & \checkmark & & & & & & & & & & & & & & & & \\
\hline
Rabbi et al.~\cite{Rabbi.2024}                              & \checkmark    & \checkmark    &               &               &               &                             & & & & & \checkmark & & & & & \checkmark & & & & & & & & & & & & \\
\hline
Torres-Arias et al.~\cite{TorresArias.2023}                 & \checkmark    &               & \checkmark    &               &               &                             & & & & & & & & & & & & & \checkmark & \checkmark & & & & & & & & \\
\hline
Tran et al.~\cite{Tran.2024}                                & \checkmark    &               &               &               &               &                             & & & & & & & & & & & & & & & & & & \checkmark & \checkmark & \checkmark & \checkmark & \\
\hline
Yu et al.~\cite{Yu.2024}                                    & \checkmark    &               & \checkmark    & \checkmark    & \checkmark    &                             & & & & & & & & & & & & & & & & & & & & & & \\
\hline
Wang et al.~\cite{Wang.2026b} & \checkmark & \checkmark & & \checkmark & & & & & & & & & & & & \checkmark & \checkmark & & & & & & & & & & & \checkmark \\
\hline
Kong et al.~\cite{Kong.2025} & & \checkmark & & \checkmark & & \checkmark & & & &  & & & & & \checkmark & & & & & & & & & & & & & \\
\hline
\end{tabular}%
}
\caption{Comparison of papers and the evaluated SBOM tools.}
\label{tab:sbom-tools}
\end{table*}

\end{document}